\documentclass[fleqn,usenatbib]{mnras}

\usepackage[T1]{fontenc}

\usepackage{graphicx}	% Including figure files
\usepackage{amsmath}	% Advanced maths commands
\usepackage{amssymb}	% Extra maths symbols

\usepackage{float}

\title[SMART]
{SMART: spectral energy distributions Markov chain analysis with radiative transfer models}

\author[Charalambia Varnava]{\parbox{\linewidth}{Charalambia Varnava\thanks{E-mail: \href{mailto:varnava.haris@gmail.com}{varnava.haris@gmail.com} (CV)} and
Andreas Efstathiou\thanks{E-mail: \href{mailto:a.efstathiou@euc.ac.cy}{a.efstathiou@euc.ac.cy} (AE)}
}
\\
\\
School of Sciences, European University Cyprus, Diogenes street, Engomi, 1516 Nicosia, Cyprus\\
}

\vspace{-40pt}

\date{Accepted 2024 April 25; Received 2024 April 24; in original form 2023 October 27}

\pubyear{2024}

\begin{document}

\pagerange{\pageref{firstpage}--\pageref{lastpage}} \pubyear{2024}

\maketitle

\label{firstpage}

\begin{abstract}
In this paper we present the publicly available open-source spectral energy distribution (SED) fitting code SMART (Spectral energy distributions Markov chain Analysis with Radiative Transfer models). Implementing a Bayesian Markov chain Monte Carlo (MCMC) method, SMART fits the ultraviolet to millimetre SEDs of galaxies exclusively with radiative transfer models that currently constitute four types of pre-computed libraries, which describe the starburst, active galactic nucleus (AGN) torus, host galaxy and polar dust components. An important novelty of SMART is that, although it fits SEDs exclusively with radiative transfer models, it takes comparable time to popular energy balance methods to run. Here we describe the key features of SMART and test it by fitting the multi-wavelength SEDs of the 42 local ultraluminous infrared galaxies (ULIRGs) that constitute the HERschel Ultraluminous Infrared Galaxy Survey (HERUS) sample. The Spitzer spectroscopy data of the HERUS ULIRGs are included in the fitting at a spectral resolution, which is matched to that of the radiative transfer models. We also present other results that highlight the performance and versatility of SMART. SMART promises to be a useful tool for studying galaxy evolution in the \textit{JWST} era. SMART is developed in PYTHON and is available at \url{https://github.com/ch-var/SMART.git}.
\end{abstract}


\vspace{-50pt}

\section{Introduction}\label{intro}

One of the most important and challenging areas of research in modern Astrophysics concerns the study of the sequence of events that led to the formation of galaxies and the supermassive black holes (SMBHs) that usually reside at their centres. Understanding of the complex astrophysics that led to the observed distribution of galaxies, both in terms of properties and numbers, is usually sought within the framework of the standard $\Lambda$ cold dark matter ($\Lambda$CDM) cosmological model (e.g. \citealt{lacey16}).

It is now clear that, in order to understand the numerous processes that govern galaxy formation and evolution (star formation in secular processes and in mergers, accretion onto SMBHs and feedback associated with them), we need multi-wavelength or panchromatic observations of galaxies at all cosmic epochs. This is mainly due to the presence of cosmic dust in the interstellar medium (ISM) of galaxies, which significantly affects their ultraviolet (UV) to millimetre spectra. The necessity of characterizing the panchromatic emission of galaxies has led to a series of surveys at all wavelengths from X-ray to radio, of ever improving sensitivity, resolution and sky coverage \citep{lon03,eal10,oli12,shi19,shi21}. Most of these surveys can only be carried out from space, which has largely become possible over the last two or three decades with missions, such as IRAS, ISO, Spitzer, AKARI, Herschel, Planck, WISE, GALEX. At the same time, significant progress has been made in the development of models. Stellar population synthesis (SPS) models \citep{bru93,bru03,maraston05,conroy09,eldridge17,byrne22} are an essential ingredient of models for the emission of galaxies. Another essential ingredient is a model for interstellar dust. Such models have been developed since the mid-80s \citep{draine84,sie92,draine01,li01,li02,jones17,hensley23}.

The spectra of galaxies are usually decomposed into a number of components. It is widely acknowledged that radiative transfer models that include the effects of cosmic dust in a realistic geometry are needed for proper interpretation of the data and the self-consistent determination of a number of physical quantities of interest, such as the stellar mass of a galaxy, its current star formation rate (SFR) and active galactic nucleus (AGN) fraction. A number of radiative transfer models for the emission of starbursts, spheroidal and disc galaxies that take into account the effects of absorption, scattering and re-emission by dust have been developed \citep{mrr93b,kru94,sil98,efstathiou00,pop00,pop11,pop17,sie07,groves08,efstathiou09,camps20,efs21}. Following the development of the unified model for AGN \citep{anton85,anton93}, models for the emission of dusty tori of increasing degree of sophistication have been developed since the early-90s \citep{pier92,mrr93,efstathiou95,fritz06,sieb15,stal16,hk17}.

Model fitting techniques that can be used to aid the interpretation of multi-wavelength observations of galaxies have also been developed \citep{efs03,dac08,noll09,cheval16,eufr17,carnall18,boq19,johnson21,doore23}. However, the usual practice for the starburst and host galaxy components is to use approximate methods, such as energy balance. SED fitting results which incorporate radiative transfer models for the starburst and host galaxy component have so far been limited \citep{ver02,far02,far03,vega08,herr17,kan21,efs22}. A thorough review of spectral energy distribution (SED) fitting methods is given in \cite{pacific23}.

Our aim is to develop an SED fitting tool, which will allow fitting of the SEDs of local galaxies, for which rest frame mid-infared (MIR) spectrophotometry from Spitzer and now \textit{JWST} is available, but also for more distant galaxies, where more limited photometry is available. MIR spectrophotometry contains important features from dust, e.g. polycyclic aromatic hydrocarbon molecule (PAH) features and silicate features that are essential for constraining the energy source of galaxies \citep{efs22}. Code Investigating GALaxy Emission (CIGALE, \citealt{noll09,boq19}), being a grid-based method, only allows the specification of discrete values for the model parameters. This limits its ability to model spectrophotometry data, which are essential for decomposing the SEDs of galaxies and inferring the contribution of star formation and AGN activity. In our method we use the full range of the model parameters. This is particularly important for key parameters like the AGN torus inclination. Although CIGALE offers the option to add routines, we need a different concept to address our scientific aims. \cite{pasp21} fitted luminous infrared galaxies (LIRGs) and ultraluminous infrared galaxies (ULIRGs) with CIGALE, but they did not include the Spitzer MIR spectrophotometry. This is also the case for the studies of \cite{yang20,yang22}. We note that, of course, our method can also be applied to a broad range of high redshift galaxies with limited photometry, as we demonstrate in Section \ref{help}.

Considering all the above, we have developed a new method for fitting radiative transfer models to data, using a Markov chain Monte Carlo (MCMC) code: Spectral energy distributions Markov chain Analysis with Radiative Transfer models (SMART). We currently test SMART with a large sample of galaxies with excellent photometry and infrared (IR) spectrophotometry obtained with the Infrared Spectrograph (IRS) onboard the Spitzer Space Telescope. The novelty of SMART is that, although it utilizes exclusively radiative transfer models, it takes comparable time to popular energy balance methods.

In particular, we utilize four types of libraries of models as input to SMART for fitting the SEDs: these are the starburst \citep{efstathiou00,efstathiou09}, AGN torus \citep{efstathiou95,fritz06,sieb15,stal16}, spheroidal \citep{efs21} or disc (Efstathiou, in preparation) host and polar dust \citep{efstathiou06} models. These libraries are part of the collection of radiative transfer models named CYprus models for Galaxies and their NUclear Spectra (CYGNUS). We have also added a component of polar dust in the fitting, which has been recognized as an important component in AGN since the mid-90s (see also \citealt{efs95,efstathiou06,matt18}), and explore the use of other AGN torus models (e.g. \citealt{fritz06,sieb15,stal16}). As an MCMC sampler we utilize the publicly available \textit{emcee} code, which is a pure-PYTHON implementation of \citeauthor{goodman10}'s (2010) Affine Invariant MCMC Ensemble sampler \citep{foreman13}.

To test SMART, we fit the multi-wavelength SEDs for the HERschel Ultraluminous Infrared Galaxy Survey (HERUS) sample \citep{far13} at $z<0.27$, assembled by \cite{efs22}. We are also testing SMART with other hyperluminous obscured galaxies at $z>4$, such as the hyperluminous obscured quasar at $z\sim4.3$ discovered by \cite{efs21} in the Herschel Extragalactic Legacy Project (HELP, \citealt{shi21}).  We fit the data with our method and the results of the fitting are then post-processed to derive physical quantities, such as stellar mass, SFR, starburst timescale, AGN fraction etc., which are essential for studying galaxy evolution. In a companion paper Varnava \& Efstathiou (in preparation) carry out a detailed study with SMART of the recently discovered hyperluminous infrared galaxy (HLIRG) COS-87529 at $z=6.853$ \citep{endsley22}. Efstathiou et al. (in preparation) use SMART to put constraints on the starburst and AGN activity of heavily obscured quasars at redshifts $z\sim0.3-3$ discovered by \cite{lon15}. 

This paper is organized as follows: In Section \ref{description} we describe our approach for SED fitting, in Section \ref{HERUS} we test our method with the HERUS sample and in Section \ref{discussion} we provide case studies to display the versatility of our method. In Section \ref{conclusion} we present our conclusions, as well as a few ideas for future development. Throughout this work we assume $H_0=70$\,km\,s$^{-1}$\,Mpc$^{-1}$, $\Omega=1$ and $\Omega_{\Lambda}=0.7$.

\section{Our approach}\label{description}

SMART demonstrates the feasibility of fitting the panchromatic SEDs of a large sample of galaxies exclusively with state-of-the-art radiative transfer models and extracting the important physical parameters. \cite{efs21} and \cite{efs22}, as well as other previous studies by \cite{herr17}, \cite{matt18}, \cite{kan21}, presented results of SED fitting with radiative transfer models, using the IDL version of Spectral energy distribution Analysis Through Markov Chains (SATMC, \citealt{john13}) by employing the synthesis mode. SATMC by itself does not output physical quantities, such as stellar mass, SFRs, AGN fraction, and therefore additional post-processing routines were developed. More details of this approach are described in Efstathiou (in preparation).

In our method we address the need of developing an SED fitting code, which is specifically designed to fit efficiently galaxy SEDs, using exclusively radiative transfer models. We use the MCMC code \textit{emcee} as the core of the method. The MCMC method, although computing intensive, gives the global best fit. In addition, as the method produces the posterior parameter distribution, realistic errors on the parameters can be estimated. This will motivate further work in using similar techniques to fit the panchromatic SEDs of the large samples of galaxies that have already been assembled by projects, such as HELP.

We fit the UV to millimetre SEDs with radiative transfer models that currently constitute four types of pre-computed libraries, which describe the starburst, AGN, host galaxy and polar dust components. These four libraries are part of CYGNUS. The starburst and AGN models are assumed to be independent of redshift, whereas for the host galaxy a different library is selected depending on the redshift. SMART allows us to explore the impact of four different AGN models and therefore constrain the properties of the obscuring torus (more details are given in \citealt{efs22}, see also Section \ref{models}).

The main novel features of our approach, which make it state-of-the-art in the field of SED fitting, are:

\begin{enumerate}

\item SMART uses radiative transfer models that take properly into account the effects of dust in a realistic geometry, whereas popular energy balance methods, such as CIGALE \citep{noll09,boq19} and Multi-wavelength Analysis of Galaxy Physical Properties (MAGPHYS, \citealt{dac08}), rely on various forms of attenuation laws. SMART can uniquely model a galaxy explicitly as a spheroidal or disc galaxy, whereas it takes comparable computing time to methods like CIGALE and MAGPHYS.

\item The code is parallelized and it is considerably faster than other MCMC codes, such as SATMC.

\item We make use of four different AGN libraries, as in \cite{efs22}.

\item As in \cite{efs22}, we can also optionally add a component of polar dust in the fitting.

\item SMART is designed to fit an SED in comparable time with a spheroidal or disc host galaxy model.

\item With SMART we can fit part of a galaxy and, if necessary, switch off any of the four components (AGN torus, starburst, spheroidal/disc, polar dust).

\end{enumerate}

SMART is an open-source and publicly available SED fitting code available at \url{https://github.com/ch-var/SMART.git}. A user manual with plenty of examples is also provided on the same website.

\subsection{Description of the radiative transfer models}\label{models}

Our method relies on the availability of libraries of radiative transfer models for the emission of starburst episodes, AGN tori and host galaxies. The models are part of the collection of radiative transfer models named CYGNUS\footnote{\url{https://arc.euc.ac.cy/cygnus-project-arc}}. The parameters of the models are listed in Table \ref{tab:parameters}.

\subsubsection{Starburst models}

The starburst models currently included in CYGNUS were described in \cite{efstathiou00} and \cite{efstathiou09}. Models for massive star formation and starbursts were also presented and discussed in \cite{mrr89,mrr93b,efs94,kru94,sil98,tak03,dop05,sie07}. The starburst model assumes the following parameters, which have the range given in brackets: the age of the starburst ($t_{*}=5-35$ Myr), the giant molecular clouds' (GMCs') initial optical depth ($\tau_v=50-250$) and the time constant of the exponentially decaying SFR ($\tau_{*}=15-35$ Myr).

\subsubsection{Spheroidal host models}

The spheroidal host models are described in more detail in \cite{efs21}. The models of \cite{bru93,bru03} are used in combination with an assumed star formation history (SFH) to compute the spectrum of starlight, which is illuminating the dust throughout the model galaxy. The spectrum of starlight is assumed to be constant throughout the galaxy, but its intensity varies throughout the galaxy according to a S\'ersic profile with $n=4$, which is equivalent to de Vaucouleurs's law. \cite{efs03}  assumed an exponentially decaying SFH, whereas here we assume a delayed exponential ($\dot{M}_{\ast} \propto t \times e^{-t/\tau^s}$), where $\tau^s$ is the e-folding time of the exponential.

The spheroidal model assumes the following parameters, which have the range given in brackets: the e-folding time of the assumed delayed exponential SFH ($\tau^s=0.125-8$ Gyr), the optical depth of the galaxy from its centre to its surface ($\tau_v^s=0.1-15$) and a parameter that controls the bolometric intensity of stellar emission relative to that of the bolometric intensity of starlight in the solar neighborhood ($\psi^s=1-17$). In this paper we use two libraries of spheroidal models, which were computed assuming an age for the galaxy equal to the age of the Universe at a redshift of $z=0.1$ and 4.3. We assumed that all the stars in the galaxy formed with a Salpeter Initial Mass Function (IMF) out of gas with a metallicity of 40 per cent of solar for the $z=0.1$ library and 6.5 per cent of solar for the $z=4.3$ library.

\subsubsection{Disc host models}

The disc models library has been computed with a two-dimensional radiative transfer code that combines the features of the code of \cite{efstathiou95} for modeling AGN tori, which does not include PAHs and small grains, and the spheroidal code of \cite{efs21}, which does. In the case of the disc models the distribution of stars, dust and molecular clouds follows a double exponential (e.g. \citealt{sil98}) instead of the de Vaucouleurs’s law assumed for the spheroidal models. The disc models have as parameters the disc SFR e-folding time $\tau^d$, the starlight intensity parameter $\psi^d$ and the equatorial optical depth in the V band $\tau_{v}^d$. Additionally, the disc models have the inclination of the disc as a parameter. The same dust model is used as in the case of the spheroidal models. The library used in this paper assumes that all the stars in the galaxy formed with a Salpeter IMF out of gas with a metallicity of 20 per cent of solar. The disc models have been used in \cite{kool18} and \cite{kan21} to fit the SEDs of LIRGs. More details and applications of the disc models are given in Efstathiou (in preparation).

\subsubsection{AGN torus models}

There are a number of AGN torus models available in the literature, which make different assumptions about the obscurer geometry \citep{pier93,mrr93,gra94,efstathiou95,nen02,nen08,dull05,fritz06,hon06,sch08,hey12,sta12,stal16,efstathiou13,sieb15,hk17}. In SMART we currently explore the impact of four different AGN models, which are described in more detail in \cite{efs22}:

\begin{enumerate}

\item The CYGNUS AGN torus model.  A number of results with this combination of models using the MCMC code SATMC have previously been presented \citep{herr17,kool18,matt18,pitch19,efs21,kan21,efs22}.

\item The AGN torus model of \cite{fritz06}.

\item The two-phase AGN torus model SKIRTOR of \cite{stal16}.

\item The two-phase AGN torus model of \cite{sieb15}.

\end{enumerate}

It is generally acknowledged in the literature (e.g. \citealt{efs14,sieb15,efs22,garcia24}) that clumpy and two-phase torus models can not reproduce the deep silicate absorption features observed in objects like IRAS 08572+3915, which we discuss in Section \ref{3915} in detail.  The parameter $p$ in the SKIRTOR model takes four discrete values in the library (0, 0.5, 1 and 1.5). Assuming a value for $p<1$, leads to AGN torus spectra with shallower silicate absorption features and therefore worse fits with these models. We have therefore fixed $p$ to the value of 1. The parameter $q$ in SKIRTOR concerns the azimuthal dependence of the density distribution and it is unlikely to have much impact on the silicate features. We have fixed this value at 1. The parameters of all of the AGN torus models used in this paper are summarized in Table \ref{tab:parameters}.

\subsubsection{Polar dust model}

 The model assumes that polar dust is concentrated in discrete spherical optically thick clouds, all of which are assumed to have constant temperature for all dust grains \citep{efstathiou06}. The model assumes the same multi-grain dust mixture as in the starburst radiative transfer model \citep{efstathiou09}, but the small grains and PAHs are assumed to be destroyed by the strong radiation field of the AGN to which these clouds are directly exposed. Unlike \cite{efs22}, in SMART the temperature of the polar dust clouds $T_p$  is a free parameter in the fit and is assumed to vary in the range  800K$-$1200K. We assume that all clouds have an optical depth from the center to the surface in the V band of 100.

\subsection{Procedure for running SMART}
We have developed the PYTHON routine \textit{SMART} that makes a sequence of runs for different galaxies, which are selected from a list according to their flag. For example, in the following list we run the galaxies with flag=1001:
\begin{verbatim}
name               redshift        flag    AGN_type
IRAS00188-0856      0.12840000     1001      2
IRAS00397-1312      0.26170000     1001      2
IRAS01003-2238      0.11780000     1001      2
IRAS03158+4227      0.13440000     1001      2
IRAS03521+0028      0.15190000     1001      2
IRAS05189-2524     0.042560000     1001      2
IRAS06035-7102     0.079470000     1001      2
IRAS06206-6315     0.092440000     1001      2
IRAS07598+6508      0.14830000     1001      1
IRAS08311-2459      0.10040000     1001      2
IRAS08572+3915     0.058350000     1001      2
IRAS09022-3615     0.059640000     1001      2
IRAS10378+1109      0.13620000     1001      2
IRAS10565+2448     0.043000000     1001      2
IRAS11095-0238      0.10660000     1001      2
...
\end{verbatim} 
The first column gives the name of the galaxy, which is also the name of the file that contains the data. The second column gives the redshift of the galaxy, the third column its selection flag and the fourth column indicates whether the galaxy will be fitted with a type 1 or type 2 AGN torus model.

The fit for each galaxy takes $\sim$ 2$-$3 minutes on a machine with a few CPUs. but, because the code is parallelized, can run much faster on a supercomputer. SMART can optionally make a corner plot, in order to assess whether there are any degeneracies in the model parameters. For this plot we use the PYTHON package GETDIST \citep{lewis15}, which we incorporate into SMART.

\begin{table*}
	\centering
\caption{Parameters of the models currently used in SMART, symbols used, their assumed ranges and summary of other information about the models. In the model of Fritz et al. (2006) there are two additional parameters that define the density distribution in the radial direction ($\beta$) and azimuthal direction ($\gamma$). In this paper we assume $\beta=0$ and $\gamma=4$. In the SKIRTOR model there are two additional parameters that define the density distribution in the radial direction ($p$) and azimuthal direction ($q$). In this paper we assume $p=1$ and $q=1$. In addition, the SKIRTOR library fixes the fraction of mass inside clumps to 97 per cent. There are four additional scaling parameters for the starburst, spheroidal or disc, AGN and polar dust models, $f_{SB}$,  $f_{sph}$ or $f_{disc}$, $f_{AGN}$ and $f_p$, respectively.}
	\label{tab:parameters}
 \resizebox{\textwidth}{!}{\begin{tabular}{llll} % four columns, alignment for each
 \hline
		Parameter &  Symbol & Range &  Comments\\
		\hline
                 &  &  & \\
{\bf CYGNUS Starburst}  &  &  & \\
                 &  &  \\
Initial optical depth of GMCs & $\tau_v$  &  50$-$250  &  \cite{efstathiou00}, \cite{efstathiou09} \\
Starburst SFR e-folding time       & $\tau_{*}$  & 10$-$35 Myr  & Incorporates \cite{bru93,bru03}  \\
Starburst age      & $t_{*}$   &  5$-$35 Myr &  Metallicity=solar, Salpeter IMF \\ 
                  &            &  & Standard galactic dust mixture with PAHs\\
                  &            &  &  \\                 
                   
{\bf CYGNUS Spheroidal Host}  &  &  &  \\
                 &  &  \\
Spheroidal SFR e-folding time      & $\tau^s$  &  0.125$-$8 Gyr  & \cite{efs03}, \cite{efs21}  \\
Starlight intensity      & $\psi^s$ &  1$-$17 &  Incorporates \cite{bru93,bru03} \\ 
Optical depth     & $\tau_{v}^s$ & 0.1$-$15 &  Range of metallicities, Salpeter IMF\\ 
                  &            &  & Standard galactic dust mixture with PAHs \\
                  &            &  &  \\
                   
{\bf CYGNUS Disc Host}  &  &  &  \\
                 &  &  \\
Disc SFR e-folding time      & $\tau^d$  &  0.5$-$8 Gyr  & \cite{efs03}, Efstathiou (in preparation)  \\
Starlight intensity      & $\psi^d$ &  1$-$9 &  Incorporates \cite{bru93,bru03} \\
Optical depth     & $\tau_{v}^d$ & 0.1$-$29 &  Range of metallicities, Salpeter IMF\\
Inclination       &  $\theta_d$ & 0\degr$-$90\degr &  Standard galactic dust mixture with PAHs \\                
                  &            &  &  \\
                  &            &  &  \\                  
                               
{\bf CYGNUS AGN torus}  &  &    &  \\
                 &  &  &  \\
Torus equatorial UV optical depth   & $\tau_{uv}$  &  260$-$1490 &  Smooth tapered discs\\  
Torus ratio of outer to inner radius & $r_2/r_1$ &  20$-$100 & \cite{efstathiou95}, \cite{efstathiou13} \\   
Torus half-opening angle  & $\theta_o$  &  30\degr$-$75\degr & Standard galactic dust mixture without PAHs\\ 
Torus inclination     & $\theta_i$  &  0\degr$-$90\degr &  The subranges $\theta_o$$-$90\degr \hspace{1pt} and 0\degr$-$$\theta_o$ are assumed for\\ 
                  &            &  &  AGN\_type=2 and AGN\_type=1, respectively. \\ 
                 &            & \\
{\bf \cite{fritz06} AGN torus}  &  &  &   \\
                 &  &  & \\
Torus equatorial optical depth at 9.7$\mu m$  & $\tau_{9.7\mu m}$ &  1$-$10 & Smooth flared discs \\  
Torus ratio of outer to inner radius & $r_2/r_1$ &  10$-$150 &  \cite{fritz06}\\   
Torus half-opening angle  & $\theta_o$  &  20\degr$-$70\degr & Standard galactic dust mixture without PAHs\\ 
Torus inclination     &  $\theta_i$ &  0\degr$-$90\degr &  The subranges $\theta_o$$-$90\degr \hspace{1pt} and 0\degr$-$$\theta_o$ are assumed for\\ 
                  &            &  &  AGN\_type=2 and AGN\_type=1, respectively. \\    
                 &            & &  \\
{\bf SKIRTOR AGN torus}  &  &   &  \\
                 &  &  &  \\
Torus equatorial optical depth at 9.7$\mu m$  &  $\tau_{9.7\mu m}$ &  3$-$11 & Two-phase flared discs \\  
Torus ratio of outer to inner radius & $r_2/r_1$ &  10$-$30 &  \cite{sta12}, \cite{stal16} \\   
Torus half-opening angle  & $\theta_o$ &  20\degr$-$70\degr &  Standard galactic dust mixture without PAHs\\ 
Torus inclination     &  $\theta_i$ &  0\degr$-$90\degr & The subranges $\theta_o$$-$90\degr \hspace{1pt} and 0\degr$-$$\theta_o$ are assumed for\\ 
                  &            &  & AGN\_type=2 and AGN\_type=1, respectively. \\   
                 &            & &  \\
{\bf \cite{sieb15} AGN torus}  &  &   &  \\
                 &  &  &  \\
Cloud volume filling factor (per cent)   & $V_c$ &  1.5$-$77  & Two-phase anisotropic spheres \\  
Optical depth of the individual clouds & $A_c$  &  0$-$45 & \cite{sieb15}\\
Optical depth of the disc mid-plane & $A_d$  &  50$-$500 &  Fluffy dust mixture without PAHs\\ 
Inclination     &  $\theta_i$ &   0\degr$-$90\degr & The subranges 45\degr$-$90\degr \hspace{1pt} and 0\degr$-$45\degr are assumed for\\ 
                  &            &  &  AGN\_type=2 and AGN\_type=1, respectively. \\
                                   &            & &  \\
{\bf Polar dust}  &  &   &  \\
                 &  &  &  \\
Temperature  & $T_p$ &  800K$-$1200K  &  Optically thick spherical clouds \citep{efstathiou06} \\  
                  &            &  &   \\                     
		\hline
	\end{tabular}}
\end{table*}

A number of physical quantities are extracted with the routine \textit{post\_SMART} we have developed. The names of the physical quantities are self-explanatory, but we also list them and describe them briefly in Table \ref{tab:derived}.

\begin{table}
	\centering
\caption{Derived physical quantities for the starburst, AGN torus and host combination (spheroidal/disc), as well as the symbol used. The luminosities are integrated over 1$-$1000 $\mu m$.}
	\label{tab:derived}
\begin{tabular}{llllll} % two columns, alignment for each
		\hline
		Physical Quantity &  Symbol\\
		\hline  
Observed AGN torus Luminosity          & $L_{AGN}^{o}$\\  
Corrected AGN torus Luminosity         & $L_{AGN}^{c}$\\
Polar dust AGN Luminosity              & $L_{p}$ \\  
Starburst Luminosity                   & $L_{SB}$\\   
Spheroidal host Luminosity             & $L_{sph}$\\
Disc host Luminosity                   & $L_{disc}$\\
Total observed Luminosity              & $L_{tot}^{o}$\\  
Total corrected Luminosity             & $L_{tot}^{c}$\\   
Starburst SFR (averaged over SB age)   & $\dot{M}_*^{age}$\\
Spheroidal SFR                         & $\dot{M}_{sph}$\\
Disc SFR                               & $\dot{M}_{disc}$\\
Total SFR                              & $\dot{M}_{tot}$\\ 
Starburst Stellar Mass                 & $M^{*}_{SB}$\\
Spheroidal Stellar Mass                & $M^{*}_{sph}$\\ 
Disc Stellar Mass                      & $M^{*}_{disc}$\\ 
Total Stellar Mass                     & $M^{*}_{tot}$\\     
AGN fraction                           & $F_{AGN}$\\ 
Anisotropy correction factor           & $A$\\
		\hline
	\end{tabular}
\end{table}

\subsection{Description of the post-processing routines}
The PYTHON routine \textit{post\_SMART} post-processes the output of the MCMC code. The user has the option to post-process the data generated by \textit{SMART} for a number of galaxies, which are selected from a list according to their flag. The code gives the option to plot the residuals, which give an indication of the quality of the fit. 

We have also developed the PYTHON routine \textit{reformatter\_SMART} that reformats the output of SMART, in order to write the extracted physical quantities of the selected sample of galaxies, as well as their errors, in the form of a LaTeX table.

\subsubsection{Calculation of luminosities and AGN fraction}
SMART allows us to have the best fit spectrum for the total emission, as well as each of the four components of emission: starburst, AGN torus, host galaxy and polar dust. We calculate the total luminosity, as well as that of each component separately, by integrating the spectra over the chosen wavelength range. All luminosities given in Table \ref{tab:derived} are 1$-$1000 $\mu m$ luminosities. We can, of course, calculate the bolometric luminosities over the whole wavelength range covered by the models. Having calculated the total luminosity and that of the AGN torus, we can calculate the AGN fraction.

An important feature of the results from the SED fitting is that due to the anisotropy of the emission from the AGN torus, which is usually optically thick to its own radiation \citep{pier93,efstathiou95,efs14}, the AGN luminosity needs to be corrected by the anisotropy correction factor $A(\theta_i)$ defined in \cite{efstathiou06}:
\begin{equation}
A(\theta_i) = {{\int_0^{\pi/2} ~~S(\theta_i') ~~sin \theta_i' ~~d \theta_i' } \over {S(\theta_i) }},
\end{equation}
where $S(\theta_i)$ is the bolometric emission over the relevant wavelength range.  $A(\theta_i)$ is generally different for the IR and bolometric luminosities and is significant for all the AGN torus models considered here.

\subsubsection{Calculation of SFRs and stellar mass}
With SMART we can calculate separately the SFR and stellar mass of the starburst and the spheroidal or disc galaxy. All SFR and stellar mass estimates are calculated self-consistently, as the radiative transfer models used in the fits incorporate the SPS models of \cite{bru93,bru03}. 

\cite{efs22} discussed two different estimates for the SFR of the starburst: the SFR averaged over the age of the starburst, $\dot{M}^{age}_{*}$, where the age is determined self-consistently from the fit, and the SFR averaged over a flat timescale of 50 Myr, $\dot{M}^{50}_{*}$. \cite{efs22} derived a very good relationship for the HERUS sample between the starburst luminosity and $\dot{M}^{age}_{*}$:
\begin{equation}
\frac{{\dot{M}_*^{age}}}{M_{\odot}yr^{-1}} = (3.11\pm0.05)\times10^{-10}\frac{L_{SB}}{L_{\odot}}.
\end{equation}

In the case of the spheroidal or disc component we are interested in the SFR at the time of observation. The SFR at all times of the history of the galaxy can, of course, be obtained from the assumed SFH, which takes the form of a delayed exponential.

\subsubsection{Calculation of minimum reduced $\chi^2$}
There is significant uncertainty in the models, which is primarily due to uncertainties regarding the dust properties, the geometry, numerical errors etc. The uncertainty in the observed photometry is primarily driven by the calibration uncertainty of the instruments. We therefore define the $\chi^{2}$ statistic we use to compare the observed and the model SEDs as:
\begin{equation}\label{chi^2}
\chi^{2}=\sum_{n}\left[\frac{\left(y_{n}-\mu_{n}\right)^{2}}{(\sigma_{\mu} \ \mu_{n})^{2}+(\sigma_y \ y_{n})^{2}}\right],
\end{equation}
where $y_n$ and $\mu_n$ are the observed data and the corresponding model value respectively, $\sigma_{\mu}$ is the uncertainty of the model and $\sigma_y$ is the uncertainty of the data.

Based on our analysis, in this paper we assumed that an uncertainty of 15 per cent realistically represented our confidence in the model SEDs. We also assumed 15 per cent of uncertainty in the observed photometry. The minimum reduced $\chi^{2}$, $\chi^{2}_{min, \nu}$, can then be obtained by
$
{\chi^{2}_{min, \nu}=\frac{\chi^{2}_{min}}{\nu}},
$
where the degrees of freedom, $\nu=n-m$, equal the number of data points $n$ minus the number of fitted parameters $m$.

\section{Testing SMART with the HERUS sample}\label{HERUS}

\subsection{Sample details}
\begin{figure}
\centering
\includegraphics[width=.9\linewidth]{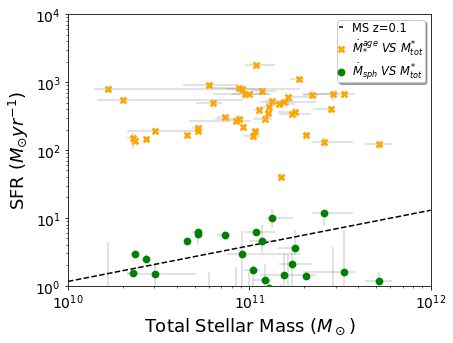}
\caption{Starburst or host SFR against total stellar mass for the HERUS sample. Also plotted is the SFR$-$stellar mass ($\dot{M}_*-M_*$) ‘main-sequence’ relation at $z=0.1$, calculated via equation (28) of Speagle et al. (2014) with t=12.161 Gyr being the age of the Universe at $z=0.1$ for the assumed cosmology.}
\label{fig:sfrVSstellar}
\end{figure}

\begin{figure*}
\centering
\includegraphics[width=.9\linewidth]{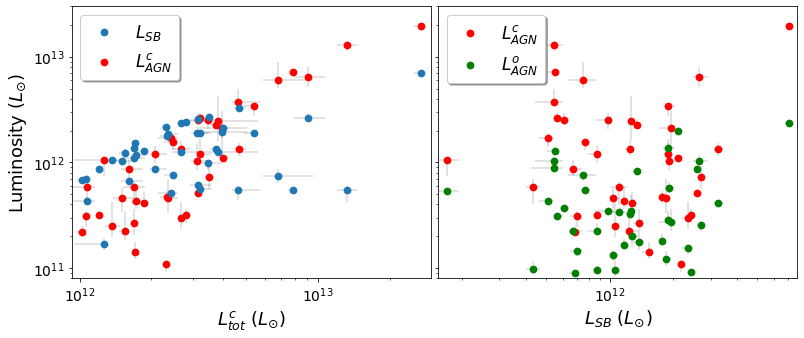}
\caption{\textit{Left:} Starburst and anisotropy-corrected AGN luminosity against total corrected IR luminosity. The starburst usually dominates the IR emission, except at $L_{tot}^c \gtrsim 3 \times 10^{12} L_{\odot}$ when AGNs start to make a significant or dominant contribution. \textit{Right:} Observed and anisotropy-corrected AGN luminosity against starburst luminosity. There is no clear evidence of a relation between $L_{SB}$ and $L_{AGN}$.}
\label{fig:SB-AGN}
\end{figure*}

The main sample of galaxies we will use for testing SMART is the HERUS sample \citep{far13}. This sample consists of 40 ULIRGs included in the Infrared Astronomical Satellite (IRAS) PSC-z survey \citep{sau00} with 60$\mu m$ fluxes greater than 2~Jy. We also include three randomly selected ULIRGs with lower 60$\mu m$ fluxes, IRAS~00397-1312 (1.8~Jy), IRAS~07598+6508 (1.7~Jy) and IRAS~13451+1232 (1.9~Jy). The quasar 3C~273 has been excluded, as it is a Blazar, to give a sample of 42 galaxies. The sample, strictly speaking, is not complete, but it includes nearly all known ULIRGs at $z<0.27$ and so provides an almost unbiased benchmark of local ULIRGs. All 42 galaxies were observed with the IRS instrument \citep{houck04} on board Spitzer and by Herschel, as part of both the HERUS and SHINING surveys \citep{fisch10,sturm11,hail12,gonz13}. HERUS studied some of the most luminous local ULIRGs, including famous objects like Arp 220, Mrk 231 and Mrk 273. In Fig. \ref{fig:sfrVSstellar} we show the plot of the SFR against the total stellar mass for the 42 galaxies of the HERUS sample. In Fig. \ref{fig:SB-AGN} we show the predicted contribution of starburst and AGN in the HERUS galaxies.

The multi-wavelength photometry and how it was assembled is described in \cite{efs22}, but we summarize the approach here as well. The far-infrared (FIR) and submillimetre photometry refers to the whole galaxy, as the aperture is always larger than the size of the galaxies. This is also the case for the IRS spectroscopy, which has a slit width $>3.6$” that corresponds to a scale $>5$kpc for a $z>0.1$ system in our sample. For the optical photometry we use the total aperture derived either from Pan-STARRS, SDSS or from \cite{u12}. The IRS data included in the fits have a wavelength grid that is separated in steps of 0.05 in the log of the rest wavelength. We also add more points around the 9.7$\mu m$ silicate feature and the PAH features to the equally spaced wavelength grid. The total number of data points we fitted varies from object to object. For example, in IRAS~03158+4227 we have a minimum of 30 data points, whereas in Arp~220 we have a maximum number of 55 data points. There is only one ULIRG in our sample with clear signs of a type 1 AGN and we discuss this object in detail in Section \ref{6508}.

\subsection{Results for the HERUS sample}\label{results}
\begin{figure}
\centering
\includegraphics[width=.9\linewidth]{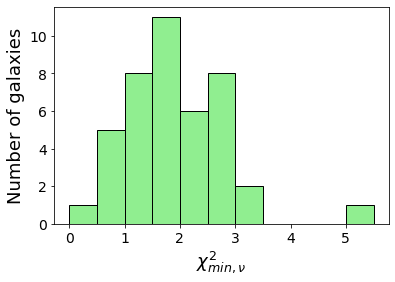}
\caption{Distribution of the reduced $\chi^2$ for the galaxies of the HERUS sample, which shows that almost all the fits have $\chi^{2}_{min, \nu} \leq 3.5$. The exception is Arp~220.}
\label{fig:histogram}
\end{figure}

\begin{figure*}
\centering
\includegraphics[width=1.\linewidth]{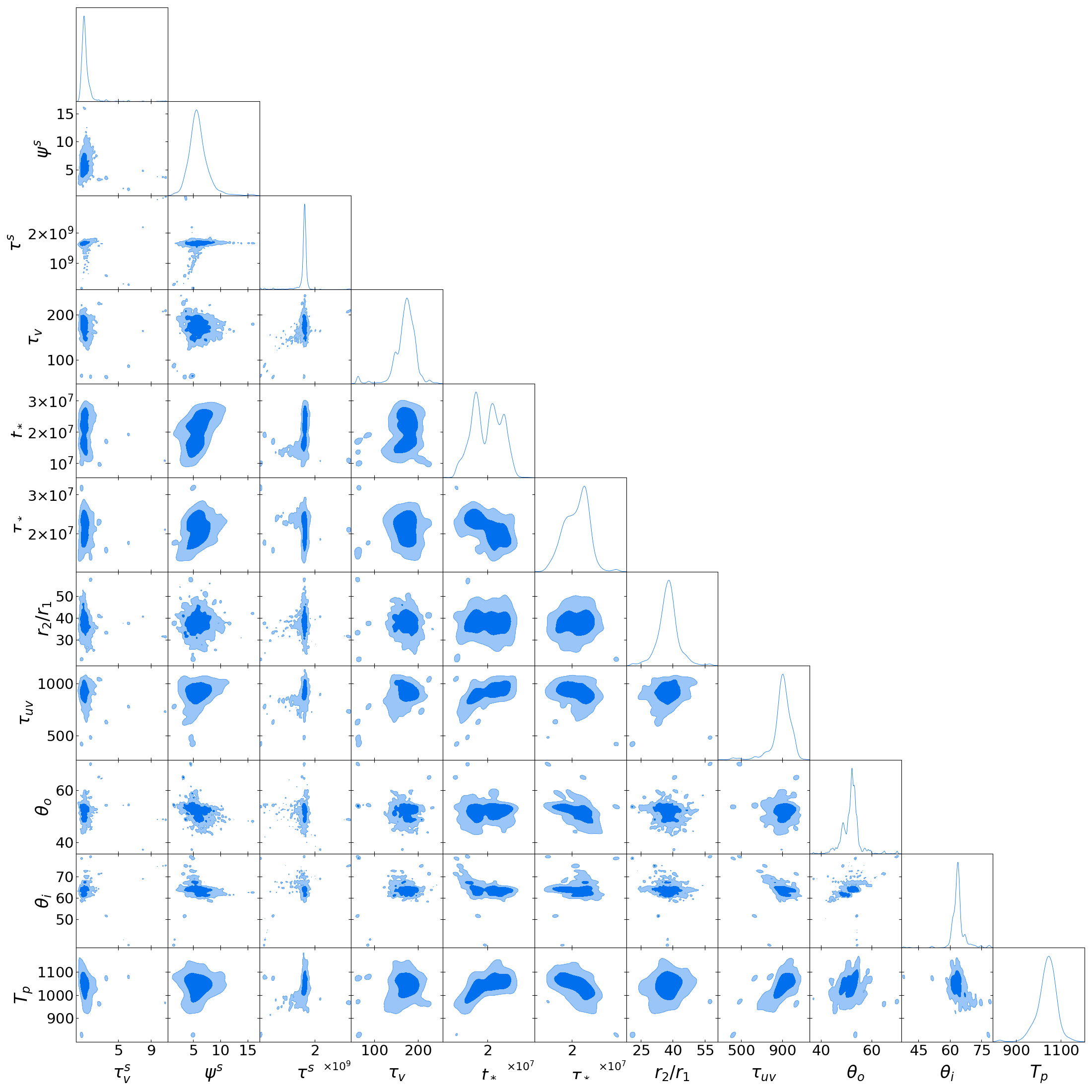}
\caption{Corner plot of the ULIRG IRAS~05189-2524, which shows that most of the parameters of the models are constrained very well with very little degeneracy}
\label{fig:corner}
\end{figure*}

In this subsection we present the results of fitting the HERUS sample with SMART. For most of the objects the fits are satisfactory and of similar quality as in the fits with SATMC reported in \cite{efs22}. Tables A1 and A2 give selected fitted parameters, along with their errors, of the HERUS sample of 42 local ULIRGs, using the CYGNUS radiative transfer models for starbursts, AGN tori, host galaxies and polar dust. In Figs A1, A2, A3, A4, A5 and A6 we present the UV to millimetre SED fits of the HERUS sample. The quality of the fits is demonstrated by a plot of the distribution of the reduced $\chi^2$ shown in Fig. \ref{fig:histogram}. We note that the distribution peaks at $\sim$ 1.75 and, apart from the fit of Arp~220 which has $\chi^{2}_{min, \nu}$=5.23, the fits for all other galaxies provide $\chi^{2}_{min, \nu}<3.16$. The values of the reduced $\chi^2$ are, of course, sensitive to the assumed values of $\sigma_\mu$ and $\sigma_y$ in equation \ref{chi^2}. Tables \ref{tab:extractedA} and \ref{tab:extractedB} give the extracted physical quantities, as well as their errors. An example of a corner plot is shown in Fig. \ref{fig:corner} for the SED fit of IRAS~05189-2524, which we discuss in more detail in Section \ref{2524}. For this corner plot we omit some nuisance parameters (our scaling parameters) to make the rest of them more visible.

\begin{figure*}
\centering
\includegraphics[width=.9\linewidth]{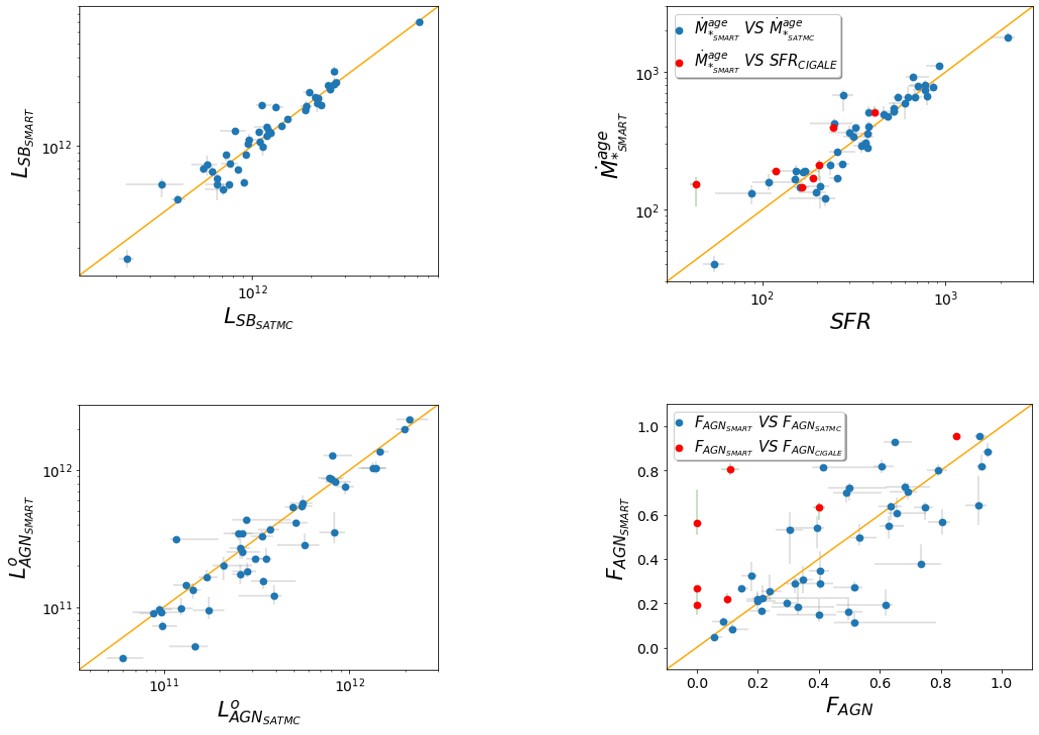}
\vspace{10pt}
\caption{Plot of representative physical quantities for the 42 HERUS local ULIRGs extracted by SMART against the same quantities extracted by Efstathiou et al. (2022) using SATMC (blue). In two of the panels we also plot with red the corresponding quantities extracted by CIGALE for seven of the HERUS ULIRGs by Paspaliaris et al. (2021).
            {\itshape Top left:} Starburst luminosity.
            {\itshape Top right:} SFR.
            {\itshape Bottom left:} Observed AGN torus luminosity.
		  {\itshape Bottom right:} AGN fraction.}
\label{fig:comparison}
\end{figure*}

\begin{table*}
	\centering
\caption{Selected extracted physical quantities for the galaxies in the HERUS sample. For the AGN and total luminosities the anisotropy-corrected luminosities are given. All of the luminosities are 1-1000 $\mu m$ luminosities.}
	\label{tab:extractedA}
 \resizebox{\textwidth}{!}{\begin{tabular}{ lcccccccc }
 \hline 
Name & $z$ & $L_{AGN}^c$ & $L_{SB}$ & $L_{sph}$ & $L_{p}$ & $L_{tot}^c$ & $\dot{M}^{age}_{*}$ & $\dot{M}_{sph}$ \\ 
     &     & $10^{12} L_\odot$ & $10^{11} L_\odot$ & $10^{10} L_\odot$ & $10^{10} L_\odot$ & $10^{12} L_\odot$ & $M_\odot yr^{-1}$ & $M_\odot yr^{-1}$ \\  
 \hline IRAS00188-0856 & 0.12840001 &    $0.43^{0.11}_{-0.1}$  &    $17.63^{0.88}_{-0.48}$  &    $3.47^{8.95}_{-0.09}$  & - &  $2.27^{0.14}_{-0.15}$  &    $592.6^{7.1}_{-69.36}$  &    $0.24^{2.71}_{-0.05}$ \\  IRAS00397-1312 & 0.2617 &    $19.42^{1.85}_{-1.52}$  &    $69.97^{5.4}_{-2.61}$  &    $5.39^{0.63}_{-0.31}$  & - &   $26.25^{1.97}_{-1.1}$  &    $1822.0^{107.8}_{-170.2}$  &    $6.17^{0.63}_{-0.8}$ \\  IRAS01003-2238 & 0.1178 &    $7.07^{0.25}_{-0.39}$  &    $5.56^{0.13}_{-0.5}$  &    $2.3^{0.14}_{-0.37}$  & - &   $7.6^{0.29}_{-0.34}$  &    $134.8^{10.72}_{-8.25}$  &    $2.94^{0.05}_{-0.44}$ \\  IRAS03158+4227 & 0.1344 &    $6.61^{1.48}_{-1.22}$  &    $26.49^{3.6}_{-1.41}$  &    $1.35^{1.71}_{-0.58}$  & - &   $9.26^{1.38}_{-1.14}$  &    $904.6^{37.53}_{-41.98}$  &    $0.42^{2.44}_{-0.42}$ \\  IRAS03521+0028 & 0.1519 &    $1.98^{0.93}_{-0.82}$  &    $19.51^{0.79}_{-2.43}$  &    $1.84^{1.25}_{-0.5}$  & - &   $3.91^{1.06}_{-0.83}$  &    $667.3^{33.28}_{-95.5}$  &    $0.0^{0.0}_{-0.0}$ \\  IRAS05189-2524 & 0.04256 &    $1.57^{0.08}_{-0.27}$  &    $7.61^{0.47}_{-0.3}$  &    $2.9^{0.17}_{-0.16}$  &    $10.76^{1.47}_{-0.85}$  &     $2.46^{0.08}_{-0.2}$  &    $191.8^{9.85}_{-9.97}$  &    $0.29^{0.03}_{-0.04}$ \\  IRAS06035-7102 & 0.07947 &    $0.4^{0.06}_{-0.06}$  &    $12.74^{0.77}_{-0.54}$  &    $9.7^{1.8}_{-5.87}$  & - &   $1.82^{0.05}_{-0.19}$  &    $342.8^{41.24}_{-32.88}$  &    $2.14^{1.73}_{-2.14}$ \\  IRAS06206-6315 & 0.09244 &    $0.5^{0.08}_{-0.11}$  &    $11.7^{0.52}_{-0.39}$  &    $8.4^{0.82}_{-2.52}$  & - &   $1.78^{0.06}_{-0.12}$  &    $396.7^{42.52}_{-41.45}$  &    $0.0^{0.01}_{-0.0}$ \\  IRAS07598+6508 & 0.1483 &    $1.11^{0.04}_{-0.07}$  &    $21.32^{0.77}_{-1.32}$  &    $0.29^{3.27}_{-0.28}$  &    $84.17^{7.89}_{-19.47}$  &     $4.02^{0.14}_{-0.16}$  &    $546.5^{39.41}_{-36.79}$  &    $0.0^{0.19}_{-0.0}$ \\  IRAS08311-2459 & 0.1004 &    $1.03^{0.46}_{-0.23}$  &    $19.02^{1.48}_{-0.81}$  &    $8.77^{0.75}_{-0.69}$  & - &   $3.08^{0.38}_{-0.15}$  &    $663.6^{70.47}_{-38.99}$  &    $1.59^{4.81}_{-1.52}$ \\  IRAS08572+3915 & 0.05835 &    $12.85^{1.1}_{-1.36}$  &    $5.46^{0.42}_{-1.07}$  &    $1.67^{0.08}_{-0.09}$  & - &   $13.34^{1.21}_{-1.4}$  &    $154.0^{17.93}_{-48.26}$  &    $1.52^{0.15}_{-0.13}$ \\  IRAS09022-3615 & 0.05964 &    $0.26^{0.13}_{-0.06}$  &    $13.75^{0.75}_{-0.63}$  &    $3.6^{0.51}_{-0.55}$  & - &   $1.69^{0.08}_{-0.05}$  &    $471.6^{12.54}_{-21.13}$  &    $0.41^{0.08}_{-0.08}$ \\  IRAS10378+1109 & 0.1362 &    $2.35^{1.57}_{-0.74}$  &    $12.68^{0.46}_{-0.54}$  &    $3.59^{0.49}_{-0.26}$  & - &  $3.68^{1.54}_{-0.81}$  &    $435.9^{25.88}_{-43.45}$  &    $1.01^{0.21}_{-0.18}$ \\  IRAS10565+2448 & 0.043 &    $0.32^{0.02}_{-0.02}$  &    $8.64^{0.32}_{-0.27}$  &    $2.07^{0.13}_{-0.08}$  & - &   $1.21^{0.03}_{-0.04}$  &    $395.6^{13.46}_{-13.02}$  &    $0.0^{0.01}_{-0.0}$ \\  IRAS11095-0238 & 0.1066 &    $6.17^{3.61}_{-0.88}$  &    $7.65^{0.84}_{-1.32}$  &    $1.72^{1.65}_{-0.36}$  & - &   $7.03^{3.35}_{-0.92}$  &    $186.3^{27.94}_{-17.52}$  &    $1.49^{0.71}_{-0.59}$ \\  IRAS12071-0444 & 0.1284 &    $2.22^{0.32}_{-0.1}$  &    $13.46^{0.06}_{-0.7}$  &    $8.4^{2.91}_{-3.0}$  & - &   $3.65^{0.11}_{-0.12}$  &    $362.8^{44.48}_{-36.01}$  &    $2.89^{3.74}_{-0.89}$ \\  IRAS13120-5453 & 0.03076 &    $0.15^{0.03}_{-0.01}$  &    $14.83^{0.77}_{-0.91}$  &    $2.85^{1.38}_{-0.66}$  & - &   $1.67^{0.06}_{-0.05}$  &    $658.9^{40.2}_{-47.42}$  &    $0.86^{1.11}_{-0.64}$ \\  IRAS13451+1232 & 0.1217 &    $3.73^{1.26}_{-0.19}$  &    $5.5^{0.5}_{-1.06}$  &    $23.45^{0.54}_{-0.42}$  &    $4.23^{2.99}_{-1.24}$  &     $4.59^{1.12}_{-0.24}$  &    $131.7^{18.56}_{-22.22}$  &    $11.79^{1.02}_{-3.94}$ \\  IRAS14348-1447 & 0.0827 &    $0.45^{0.25}_{-0.12}$  &    $18.49^{1.0}_{-0.46}$  &    $4.74^{1.5}_{-0.21}$  & - &   $2.35^{0.28}_{-0.09}$  &    $512.6^{49.4}_{-29.2}$  &    $1.67^{1.33}_{-1.6}$ \\  IRAS14378-3651 & 0.0676 &    $0.46^{0.08}_{-0.12}$  &    $10.18^{0.5}_{-0.2}$  &    $2.1^{1.88}_{-0.6}$  & - &   $1.47^{0.21}_{-0.12}$  &    $287.7^{36.94}_{-21.34}$  &    $0.02^{0.63}_{-0.01}$ \\  IRAS15250+3609 & 0.05516 &    $2.54^{0.55}_{-0.25}$  &    $5.98^{0.56}_{-0.4}$  &    $2.15^{0.33}_{-0.11}$  & - &   $3.18^{0.41}_{-0.28}$  &    $144.8^{9.01}_{-9.26}$  &    $2.48^{0.19}_{-0.1}$ \\  IRAS15462-0450 & 0.09979 &    $0.6^{0.16}_{-0.2}$  &    $11.49^{0.69}_{-1.5}$  &    $2.36^{0.31}_{-0.25}$  & - &   $1.72^{0.22}_{-0.18}$  &    $268.0^{10.33}_{-16.77}$  &    $0.73^{1.17}_{-0.71}$ \\  IRAS16090-0139 & 0.13358 &    $0.72^{0.18}_{-0.16}$  &    $27.05^{0.86}_{-0.8}$  &    $3.98^{2.82}_{-0.81}$  & - &   $3.48^{0.12}_{-0.13}$  &    $780.6^{76.4}_{-42.66}$  &    $2.56^{2.89}_{-2.56}$ \\  IRAS17208-0014 & 0.0428 &    $0.11^{0.0}_{-0.02}$  &    $21.66^{0.97}_{-0.78}$  &    $1.57^{0.1}_{-0.02}$  & - &  $2.29^{0.08}_{-0.08}$  &    $809.4^{68.56}_{-30.52}$  &    $0.19^{0.59}_{-0.08}$ \\  IRAS19254-7245 & 0.061709 &    $1.73^{0.63}_{-0.32}$  &    $5.06^{0.39}_{-0.66}$  &    $19.09^{5.14}_{-2.12}$  & - &   $2.45^{0.52}_{-0.31}$  &    $119.7^{6.24}_{-14.92}$  &    $1.08^{0.45}_{-0.3}$ \\  IRAS19297-0406 & 0.08573 &    $0.31^{0.13}_{-0.08}$  &    $23.3^{0.37}_{-1.78}$  &    $4.1^{0.11}_{-0.08}$  & - &   $2.66^{0.04}_{-0.06}$  &    $660.0^{113.1}_{-21.46}$  &    $0.0^{0.42}_{-0.0}$ \\  IRAS20087-0308 & 0.10567 &    $0.33^{0.04}_{-0.04}$  &    $23.99^{0.74}_{-1.1}$  &    $3.65^{0.78}_{-0.46}$  & - &   $2.77^{0.05}_{-0.08}$  &    $1108.0^{33.86}_{-59.3}$  &    $0.0^{0.27}_{-0.0}$ \\  IRAS20100-4156 & 0.12958 &    $1.3^{0.18}_{-0.17}$  &    $32.62^{1.63}_{-1.06}$  &    $6.27^{5.04}_{-0.9}$  & - &   $4.65^{0.26}_{-0.25}$  &    $763.1^{43.59}_{-50.99}$  &    $4.56^{3.76}_{-0.94}$ \\  IRAS20414-1651 & 0.087084 &    $0.22^{0.12}_{-0.04}$  &    $12.35^{1.23}_{-0.23}$  &    $1.31^{0.04}_{-0.07}$  & - &   $1.56^{0.01}_{-0.09}$  &    $495.1^{72.17}_{-4.58}$  &    $0.18^{0.34}_{-0.07}$ \\  IRAS20551-4250 & 0.042996 &    $0.84^{0.18}_{-0.11}$  &    $6.65^{0.63}_{-0.43}$  &    $4.85^{0.73}_{-0.41}$  & - &   $1.56^{0.18}_{-0.13}$  &    $163.6^{20.84}_{-11.2}$  &    $4.64^{0.38}_{-0.4}$ \\  IRAS22491-1808 & 0.0778 &    $1.13^{1.15}_{-0.17}$  &    $8.39^{0.72}_{-1.2}$  &    $7.9^{0.3}_{-4.03}$  & - &   $2.07^{1.13}_{-0.2}$  &    $209.5^{21.62}_{-46.27}$  &    $6.27^{0.29}_{-2.25}$ \\  IRAS23128-5919 & 0.0446 &    $0.3^{0.04}_{-0.04}$  &    $6.95^{0.39}_{-0.43}$  &    $5.48^{1.56}_{-0.66}$  & - &   $1.06^{0.03}_{-0.06}$  &    $187.5^{8.98}_{-17.01}$  &    $5.61^{0.74}_{-0.88}$ \\  IRAS23230-6926 & 0.10659 &    $1.33^{0.27}_{-0.15}$  &    $12.28^{0.56}_{-0.81}$  &    $9.41^{0.57}_{-1.79}$  & - &   $2.62^{0.27}_{-0.13}$  &    $303.4^{10.07}_{-14.64}$  &    $5.74^{0.56}_{-0.96}$ \\  IRAS23253-5415 & 0.13 &    $1.16^{0.2}_{-0.07}$  &    $19.08^{0.67}_{-0.95}$  &    $6.83^{0.74}_{-0.83}$  & - &   $3.18^{0.09}_{-0.11}$  &    $674.3^{52.35}_{-80.31}$  &    $0.82^{3.14}_{-0.79}$ \\  IRAS23365+3604 & 0.0645 &    $2.63^{0.37}_{-0.54}$  &    $9.84^{1.34}_{-1.28}$  &    $5.64^{1.21}_{-2.14}$  & - &   $3.61^{0.54}_{-0.47}$  &    $285.7^{101.3}_{-21.5}$  &    $1.22^{1.32}_{-0.73}$ \\  Mrk1014 & 0.16311 &    $0.51^{0.02}_{-0.04}$  &    $25.87^{0.99}_{-1.4}$  &    $0.01^{6.39}_{-0.01}$  & - &   $3.12^{0.09}_{-0.13}$  &    $787.1^{49.55}_{-62.5}$  &    $0.0^{4.95}_{-0.0}$ \\  UGC5101 & 0.039367 &    $0.22^{0.04}_{-0.02}$  &    $6.85^{0.31}_{-0.24}$  &    $11.25^{0.36}_{-0.45}$  & - &   $1.02^{0.03}_{-0.03}$  &    $168.6^{9.03}_{-8.99}$  &    $1.43^{0.08}_{-0.14}$ \\  Mrk231 & 0.04217 &    $3.4^{0.38}_{-0.62}$  &    $18.75^{1.6}_{-0.65}$  &    $10.23^{3.05}_{-1.23}$  & - &   $5.35^{0.44}_{-0.37}$  &    $515.5^{33.35}_{-37.06}$  &    $10.58^{2.95}_{-3.16}$ \\  Mrk273 & 0.03778 &    $0.26^{0.13}_{-0.04}$  &    $10.54^{0.43}_{-0.44}$  &    $3.52^{0.29}_{-0.25}$  & - &   $1.38^{0.1}_{-0.08}$  &    $356.0^{17.54}_{-14.23}$  &    $0.46^{0.1}_{-0.1}$ \\  Mrk463 & 0.050355 &    $1.05^{0.05}_{-0.3}$  &    $1.7^{0.24}_{-0.23}$  &    $3.78^{0.83}_{-0.55}$  & - &   $1.26^{0.02}_{-0.32}$  &    $40.28^{5.74}_{-5.5}$  &    $0.05^{1.01}_{-0.04}$ \\  Arp220 & 0.018 &    $2.6^{0.34}_{-0.29}$  &    $5.71^{0.3}_{-0.22}$  &    $1.93^{0.04}_{-0.04}$  & - &   $3.2^{0.38}_{-0.26}$  &    $216.8^{4.27}_{-12.76}$  &    $0.03^{0.01}_{-0.0}$ \\  NGC6240 & 0.0244 &    $0.59^{0.13}_{-0.19}$  &    $4.33^{0.3}_{-0.29}$  &    $4.89^{0.78}_{-1.06}$  & - &   $1.08^{0.15}_{-0.17}$  &    $162.5^{19.42}_{-22.86}$  &    $1.48^{0.44}_{-0.65}$ \\   \hline 
 \end{tabular}}
\end{table*}

\begin{table*}
	\centering
\caption{Other extracted physical quantities for the galaxies in the HERUS sample}
	\label{tab:extractedB}
 \resizebox{\textwidth}{!}{\begin{tabular}{ lcccccccc }
 \hline 
Name & $\dot{M}_{tot}$ & ${M}^{*}_{sph}$ & ${M}^{*}_{SB}$ & ${M}^{*}_{tot}$& $F_{AGN}$ & $A$ \\ 
     & $M_\odot yr^{-1}$ & $10^{10} M_\odot$ & $10^{9} M_\odot$ & $10^{10} M_\odot$ & $   $ & $   $ \\  
 \hline IRAS00188-0856 &    $592.8^{7.29}_{-65.63}$  &    $14.44^{2.94}_{-0.13}$  &    $17.25^{0.03}_{-3.02}$  &    $16.16^{2.64}_{-0.2}$  &     $0.19^{0.03}_{-0.04}$  &     $2.48^{0.37}_{-0.26}$ \\  IRAS00397-1312 &    $1827.0^{109.7}_{-169.2}$  &    $9.7^{2.66}_{-1.79}$  &    $13.19^{1.34}_{-1.91}$  &    $11.03^{2.79}_{-1.72}$  &     $0.73^{0.03}_{-0.02}$  &     $8.26^{0.24}_{-0.3}$ \\  IRAS01003-2238 &    $137.6^{10.41}_{-8.06}$  &    $2.21^{0.06}_{-0.27}$  &    $1.74^{1.11}_{-0.25}$  &    $2.35^{0.09}_{-0.13}$  &     $0.93^{0.0}_{-0.01}$  &     $5.57^{0.1}_{-0.06}$ \\  IRAS03158+4227 &    $907.5^{34.95}_{-44.92}$  &    $4.0^{3.42}_{-1.96}$  &    $24.02^{2.03}_{-2.02}$  &    $6.31^{3.59}_{-1.74}$  &     $0.7^{0.05}_{-0.04}$  &     $6.43^{0.8}_{-1.01}$ \\  IRAS03521+0028 &    $667.3^{33.28}_{-95.5}$  &    $7.96^{3.06}_{-1.55}$  &    $19.9^{1.03}_{-2.48}$  &    $9.94^{3.17}_{-1.44}$  &     $0.51^{0.08}_{-0.13}$  &     $7.67^{0.87}_{-2.51}$ \\  IRAS05189-2524 &    $192.2^{9.78}_{-10.01}$  &    $10.42^{0.55}_{-0.53}$  &    $3.79^{0.76}_{-0.56}$  &    $10.79^{0.64}_{-0.55}$  &     $0.63^{0.02}_{-0.06}$  &     $2.89^{0.08}_{-0.06}$ \\  IRAS06035-7102 &    $344.6^{39.43}_{-30.82}$  &    $16.99^{4.01}_{-2.43}$  &    $2.68^{7.38}_{-0.45}$  &    $17.41^{4.51}_{-2.6}$  &     $0.22^{0.03}_{-0.02}$  &     $2.13^{0.34}_{-0.13}$ \\  IRAS06206-6315 &    $396.7^{42.52}_{-41.45}$  &    $26.0^{2.22}_{-3.49}$  &    $10.83^{2.07}_{-1.05}$  &    $27.07^{1.88}_{-3.1}$  &     $0.28^{0.04}_{-0.05}$  &     $2.88^{0.45}_{-0.4}$ \\  IRAS07598+6508 &    $546.8^{39.19}_{-36.99}$  &    $0.96^{8.73}_{-0.92}$  &    $13.51^{1.63}_{-1.87}$  &    $2.33^{8.79}_{-0.86}$  &     $0.28^{0.02}_{-0.02}$  &     $0.55^{0.02}_{-0.01}$ \\  IRAS08311-2459 &    $664.2^{74.29}_{-37.32}$  &    $31.8^{4.63}_{-13.73}$  &    $17.2^{1.5}_{-0.85}$  &    $33.47^{4.69}_{-13.63}$  &     $0.34^{0.09}_{-0.06}$  &     $1.79^{0.54}_{-0.25}$ \\  IRAS08572+3915 &    $155.5^{18.02}_{-48.19}$  &    $2.19^{0.19}_{-0.15}$  &    $0.9^{0.14}_{-0.07}$  &    $2.28^{0.22}_{-0.16}$  &     $0.96^{0.01}_{-0.01}$  &     $14.2^{1.38}_{-1.29}$ \\  IRAS09022-3615 &    $471.9^{12.67}_{-21.17}$  &    $12.93^{2.89}_{-1.95}$  &    $13.12^{0.8}_{-0.81}$  &    $14.3^{2.8}_{-1.94}$  &     $0.15^{0.06}_{-0.03}$  &     $1.53^{0.34}_{-0.16}$ \\  IRAS10378+1109 &    $436.9^{25.77}_{-40.16}$  &    $11.46^{0.71}_{-1.38}$  &    $11.62^{0.61}_{-1.06}$  &    $12.68^{0.66}_{-1.47}$  &     $0.64^{0.11}_{-0.09}$  &     $6.07^{1.94}_{-0.5}$ \\  IRAS10565+2448 &    $395.6^{13.45}_{-13.02}$  &    $9.94^{0.7}_{-0.51}$  &    $12.13^{0.39}_{-0.48}$  &    $11.13^{0.67}_{-0.5}$  &     $0.27^{0.01}_{-0.02}$  &     $3.31^{0.09}_{-0.26}$ \\  IRAS11095-0238 &    $188.3^{26.93}_{-18.01}$  &    $2.5^{1.28}_{-0.82}$  &    $3.83^{1.23}_{-1.64}$  &    $2.79^{1.38}_{-0.64}$  &     $0.88^{0.06}_{-0.02}$  &     $8.1^{3.3}_{-0.79}$ \\  IRAS12071-0444 &    $365.7^{44.46}_{-36.9}$  &    $17.75^{0.7}_{-3.39}$  &    $2.17^{1.0}_{-0.21}$  &    $17.97^{0.68}_{-3.37}$  &     $0.61^{0.07}_{-0.01}$  &     $2.82^{0.39}_{-0.16}$ \\  IRAS13120-5453 &    $659.2^{40.75}_{-44.06}$  &    $6.68^{2.93}_{-2.56}$  &    $19.86^{1.14}_{-2.21}$  &    $8.74^{2.9}_{-2.85}$  &     $0.09^{0.02}_{-0.01}$  &     $2.1^{0.19}_{-0.13}$ \\  IRAS13451+1232 &    $143.8^{20.13}_{-27.22}$  &    $25.61^{11.43}_{-3.77}$  &    $1.8^{0.1}_{-0.34}$  &    $25.8^{11.38}_{-3.76}$  &     $0.82^{0.05}_{-0.01}$  &     $3.65^{0.71}_{-0.11}$ \\  IRAS14348-1447 &    $514.3^{50.25}_{-29.42}$  &    $13.48^{8.48}_{-3.64}$  &    $15.4^{1.18}_{-0.82}$  &    $14.98^{8.51}_{-3.61}$  &     $0.19^{0.07}_{-0.04}$  &     $3.73^{1.3}_{-0.47}$ \\  IRAS14378-3651 &    $289.2^{35.42}_{-22.87}$  &    $8.1^{0.77}_{-1.64}$  &    $8.91^{1.01}_{-0.75}$  &    $8.95^{0.76}_{-1.44}$  &     $0.31^{0.02}_{-0.06}$  &     $3.6^{0.23}_{-0.68}$ \\  IRAS15250+3609 &    $147.5^{9.02}_{-9.49}$  &    $2.32^{0.1}_{-0.11}$  &    $3.81^{0.45}_{-0.49}$  &    $2.68^{0.11}_{-0.09}$  &     $0.81^{0.03}_{-0.03}$  &     $6.9^{0.65}_{-0.53}$ \\  IRAS15462-0450 &    $269.0^{10.0}_{-17.79}$  &    $8.54^{1.96}_{-4.37}$  &    $4.4^{1.09}_{-1.18}$  &    $9.03^{2.08}_{-4.38}$  &     $0.34^{0.06}_{-0.06}$  &     $1.73^{0.46}_{-0.46}$ \\  IRAS16090-0139 &    $783.4^{77.66}_{-45.3}$  &    $8.84^{8.21}_{-3.58}$  &    $23.86^{2.13}_{-1.45}$  &    $11.21^{8.18}_{-3.64}$  &     $0.21^{0.04}_{-0.04}$  &     $2.73^{0.5}_{-0.31}$ \\  IRAS17208-0014 &    $809.6^{69.15}_{-30.6}$  &    $6.31^{0.03}_{-2.71}$  &    $24.22^{2.55}_{-0.47}$  &    $8.72^{0.01}_{-2.44}$  &     $0.05^{0.0}_{-0.01}$  &     $2.52^{0.27}_{-0.37}$ \\  IRAS19254-7245 &    $121.0^{6.34}_{-15.08}$  &    $51.24^{12.16}_{-7.97}$  &    $1.25^{0.14}_{-0.13}$  &    $51.38^{12.17}_{-7.98}$  &     $0.71^{0.07}_{-0.05}$  &     $3.91^{0.55}_{-0.36}$ \\  IRAS19297-0406 &    $660.0^{113.5}_{-21.24}$  &    $20.25^{1.96}_{-2.62}$  &    $20.3^{2.95}_{-0.89}$  &    $22.26^{2.04}_{-2.58}$  &     $0.12^{0.05}_{-0.03}$  &     $2.0^{0.56}_{-0.28}$ \\  IRAS20087-0308 &    $1108.0^{33.56}_{-58.6}$  &    $15.61^{2.0}_{-2.17}$  &    $33.77^{1.05}_{-1.85}$  &    $18.99^{2.13}_{-2.11}$  &     $0.12^{0.02}_{-0.02}$  &     $3.51^{0.36}_{-0.34}$ \\  IRAS20100-4156 &    $768.0^{42.35}_{-50.45}$  &    $10.47^{2.3}_{-1.39}$  &    $12.98^{3.04}_{-4.55}$  &    $11.95^{2.12}_{-1.66}$  &     $0.28^{0.03}_{-0.02}$  &     $3.16^{0.41}_{-0.24}$ \\  IRAS20414-1651 &    $495.6^{71.76}_{-4.92}$  &    $5.25^{0.15}_{-1.7}$  &    $15.02^{1.83}_{-0.06}$  &    $6.74^{0.34}_{-1.69}$  &     $0.15^{0.07}_{-0.03}$  &     $4.46^{2.07}_{-0.7}$ \\  IRAS20551-4250 &    $168.1^{20.4}_{-10.63}$  &    $4.31^{0.23}_{-0.16}$  &    $2.07^{0.66}_{-0.49}$  &    $4.55^{0.15}_{-0.14}$  &     $0.55^{0.03}_{-0.06}$  &     $3.84^{0.45}_{-0.38}$ \\  IRAS22491-1808 &    $215.7^{21.48}_{-46.05}$  &    $4.98^{0.14}_{-0.99}$  &    $2.59^{4.17}_{-0.83}$  &    $5.21^{0.14}_{-0.51}$  &     $0.56^{0.15}_{-0.06}$  &     $5.01^{1.68}_{-0.42}$ \\  IRAS23128-5919 &    $192.8^{9.28}_{-17.54}$  &    $4.84^{0.29}_{-0.21}$  &    $2.54^{0.69}_{-0.84}$  &    $5.16^{0.19}_{-0.18}$  &     $0.29^{0.02}_{-0.04}$  &     $2.14^{0.14}_{-0.16}$ \\  IRAS23230-6926 &    $308.5^{11.22}_{-14.01}$  &    $6.9^{0.53}_{-0.45}$  &    $3.72^{1.0}_{-0.76}$  &    $7.28^{0.54}_{-0.44}$  &     $0.51^{0.04}_{-0.04}$  &     $4.08^{0.39}_{-0.4}$ \\  IRAS23253-5415 &    $674.9^{52.35}_{-79.95}$  &    $27.38^{1.65}_{-8.61}$  &    $19.42^{0.95}_{-2.51}$  &    $29.73^{1.41}_{-9.01}$  &     $0.37^{0.04}_{-0.02}$  &     $4.22^{0.43}_{-0.29}$ \\  IRAS23365+3604 &    $288.3^{99.23}_{-22.81}$  &    $11.77^{0.71}_{-2.34}$  &    $6.68^{1.97}_{-1.61}$  &    $12.27^{1.0}_{-2.18}$  &     $0.72^{0.02}_{-0.06}$  &     $7.42^{2.17}_{-1.36}$ \\  Mrk1014 &    $787.1^{49.55}_{-58.96}$  &    $0.01^{14.36}_{-0.01}$  &    $15.0^{1.76}_{-2.19}$  &    $1.87^{13.85}_{-0.39}$  &     $0.16^{0.0}_{-0.01}$  &     $0.6^{0.01}_{-0.03}$ \\  UGC5101 &    $169.8^{9.08}_{-8.86}$  &    $20.14^{1.26}_{-0.77}$  &    $1.66^{0.14}_{-0.15}$  &    $20.32^{1.24}_{-0.78}$  &     $0.22^{0.03}_{-0.02}$  &     $2.42^{0.24}_{-0.15}$ \\  Mrk231 &    $524.3^{35.12}_{-36.04}$  &    $12.39^{3.3}_{-0.74}$  &    $10.03^{1.92}_{-2.49}$  &    $13.32^{3.4}_{-0.86}$  &     $0.64^{0.02}_{-0.06}$  &     $2.44^{0.2}_{-0.34}$ \\  Mrk273 &    $356.3^{17.61}_{-13.92}$  &    $11.77^{0.66}_{-0.47}$  &    $9.6^{0.75}_{-0.71}$  &    $12.7^{0.65}_{-0.42}$  &     $0.19^{0.08}_{-0.03}$  &     $2.68^{0.73}_{-0.35}$ \\  Mrk463 &    $40.28^{6.02}_{-5.33}$  &    $14.9^{0.82}_{-0.82}$  &    $0.62^{0.31}_{-0.22}$  &    $14.99^{0.78}_{-0.85}$  &     $0.82^{0.03}_{-0.03}$  &     $1.97^{0.05}_{-0.57}$ \\  Arp220 &    $216.8^{4.28}_{-12.76}$  &    $8.71^{0.12}_{-0.27}$  &    $5.66^{0.2}_{-0.3}$  &    $9.24^{0.16}_{-0.24}$  &     $0.82^{0.01}_{-0.03}$  &     $8.53^{0.52}_{-0.28}$ \\  NGC6240 &    $163.6^{19.99}_{-21.91}$  &    $10.08^{0.96}_{-1.11}$  &    $4.84^{0.44}_{-1.06}$  &    $10.52^{0.97}_{-1.12}$  &     $0.54^{0.06}_{-0.1}$  &     $5.73^{0.84}_{-0.78}$ \\   \hline 
 \end{tabular}}
\end{table*}

There are two ULIRGS (IRAS~01003-2238 and Mrk 1014) for which the fit is unsatisfactory, despite repeated runs of the code. As also noted by \cite{efs22}, in these objects there is most probably a dual AGN, one viewed with its torus edge-on and dominating in the MIR, and one viewed face-on, which dominates in the optical/UV. We have not developed yet SMART to deal with the case of dual AGN, so in Figs A1 and A5 we plot the fit with a single AGN.

For three of the ULIRGs plotted in Figs A1, A2 and A3 we run the code with polar dust enabled, as we see evidence that the addition of polar dust can improve the fit. These are IRAS~05189-2524, IRAS~07598+6508 and IRAS~13451+1232. We discuss IRAS~05189-2524 and IRAS~07598+6508 in more detail in Sections \ref{2524} and \ref{6508}, respectively.

\subsection{Comparison with results from other approaches}

In Fig. \ref{fig:comparison} we compare key physical quantities derived by SMART with those obtained with SATMC, using the same multi-component CYGNUS radiative transfer models as in \cite{efs22}. Regarding the luminosity and the SFR of the starburst, we see that there is very good agreement between the estimates of SMART and SATMC, with only a few exceptions. Some of the exceptions may be attributed to the statistical nature of the methods and this issue certainly deserves further exploration to understand these differences better. The corresponding comparison for the observed AGN torus luminosity shows that there is generally good agreement between the estimates with SMART and SATMC. For the AGN fraction, as expected, there is more discrepancy. We attribute the more scatter in the AGN fraction to the fact that it incorporates the anisotropy corrections, so it is much more sensitive to changes to best fit parameters compared to other quantities, such as the luminosity of the starburst, the SFR of the starburst and the observed AGN torus luminosity. We conclude that for most of the galaxies the fits with SMART are satisfactory and of similar quality as in the fits with SATMC reported in \cite{efs22}. It is, however, important to note that, even using the same models, significant differences can arise in key physical quantities due to the statistical nature of the SED fitting methods.

Eleven of the sources in the sample of \cite{pasp21} are ULIRGs and seven of them are included in the HERUS sample. We compare the results of \cite{pasp21} obtained with CIGALE and using the SKIRTOR AGN torus model with those obtained with our method in Fig. \ref{fig:comparison}. We observe good agreement in the SFR predicted by SMART and CIGALE. The only exception is IRAS~08572+3915 for which CIGALE predicts a much lower SFR. This is the galaxy we discuss in detail in Section \ref{3915}. The model of \cite{pasp21} fails to reproduce the deep silicate absorption feature of this ULIRG, which in our model is attributed to the fact that the smooth AGN torus is viewed almost edge-on. It is well understood that two-phase models like SKIRTOR can not produce deep silicate absorption features like the one observed in IRAS~08572+3915 (see, for example, discussion in \citealt{efs22}). The results of \cite{pasp21} do not take into account the anisotropy correction and this is probably one of the main reasons the AGN fraction of the seven ULIRGs in common with our sample are lower.

\section{Discussion}\label{discussion}
In this section we discuss the versatility of SMART by providing a number of case studies. As part of this demonstration, we present a set of SED fits derived from SMART and highlighting the key advantages of the method. All of the following SED fits are done with the CYGNUS multi-component radiative transfer models. In Sections \ref{3915} and \ref{help} we present a comparison of our results for the galaxies IRAS~08572+3915 and HELP\_J100156.75+022344.7, respectively, with the other AGN models that we have available (the smooth torus model of \citealt{fritz06} and the two-phase models of \citealt{sieb15} and SKIRTOR).

\subsection{Fit of a ULIRG associated with polar dust}\label{2524}
 IRAS~05189-2524 is generally believed to be a `warm' ULIRG and to contain an obscured AGN with its torus viewed almost edge-on. This galaxy was recognized from the mid-90s as a ULIRG that has similar characteristics to the prototypical Seyfert 2 galaxy NGC~1068. It was found in particular by \cite{young96} to show broad lines in polarized flux. NGC~1068 was also the first LIRG that was recognized to have polar dust \citep{braatz93,cam93,efs95}. \cite{efs22} found that a fit with polar dust provides a better fit to the SED of IRAS~05189-2524 compared to a fit without polar dust. \cite{reyn22} found evidence that IRAS~05189-2524 recently experienced a dust obscured tidal disruption event (TDE) with an IR echo from polar dust. This phenomenon was first proposed for the TDE in the LIRG Arp~299 by \cite{matt18}.
 
 We find that a fit with SMART with an obscured AGN and polar dust provides a good fit to the SED. The model for polar dust incorporated in SMART has the temperature of polar dust as a free parameter, whereas \cite{efs22} considered only a single temperature for the polar dust. We find that the polar dust in IRAS~05189-2524 has a temperature of $1050.6^{28.7}_{-23.2}$K. In Tables A1, A2, \ref{tab:extractedA} and \ref{tab:extractedB} we list all of the important fitted parameters and extracted physical properties for this ULIRG. The SED fit plot is shown in Fig. A1.

\subsection{Fit of a ULIRG with an unobscured quasar and also associated with polar dust}\label{6508}
The galaxy IRAS~07598+6508 is the only ULIRG in the HERUS sample (apart from 3C~273 which we do not fit, as it clearly has a strong contribution in the submillimetre from radio emission), which shows clear signs of a type 1 AGN with a strong optical continuum with a power-law shape. This ULIRG  was therefore fitted with a type 1 AGN by \cite{efs22}.

\begin{figure*}
\centering
\includegraphics[width=.9\linewidth]{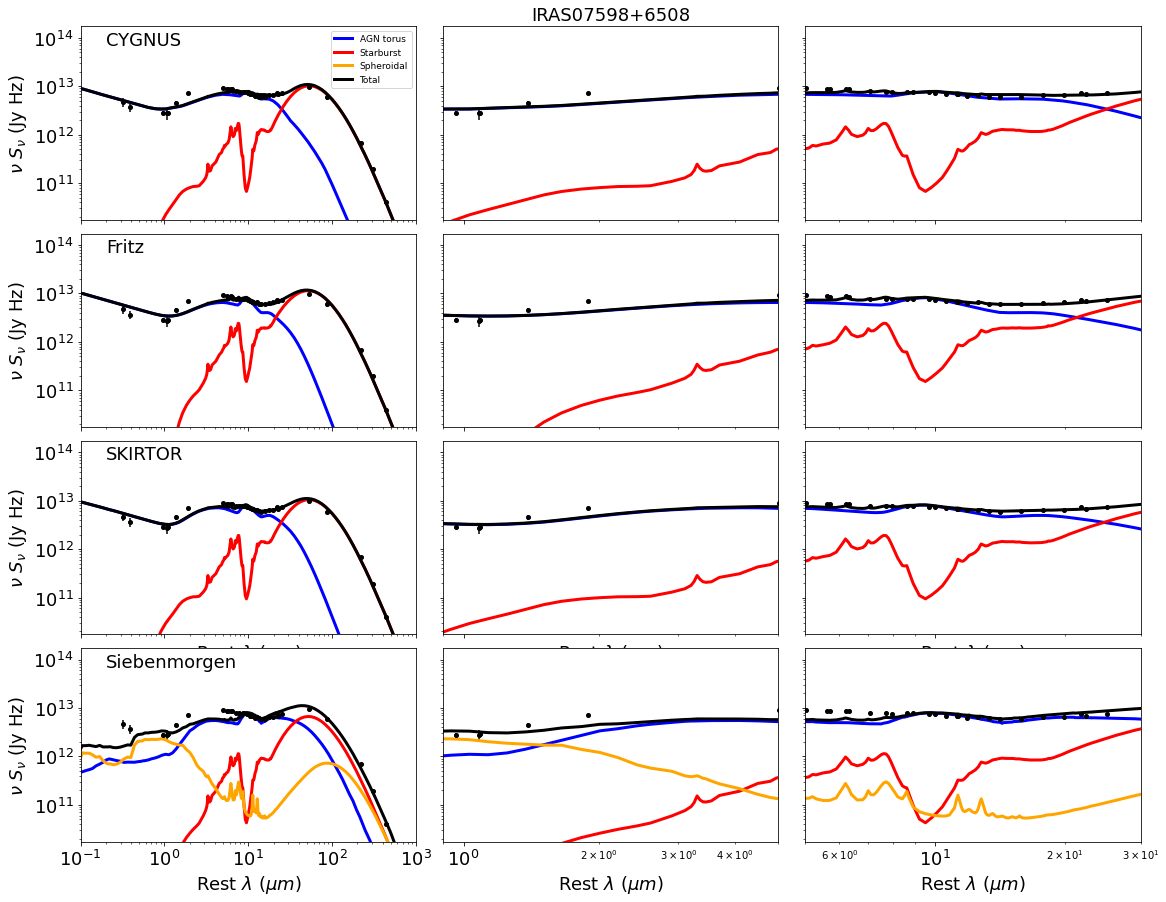}
\caption{Comparison SED fit plots of IRAS~07598+6508 without using the polar dust component. The AGN torus, starburst, spheroidal host and total emission
are plotted as shown in the legend. The top row shows fits with the CYGNUS combination of models. The second row shows fits with the CYGNUS AGN model replaced by the model of Fritz et al. (2006), the third row replaces the CYGNUS AGN model with the SKIRTOR model, while the bottom row replaces the CYGNUS AGN model with the model of Siebenmorgen et al. (2015). The left panel shows the fit over 0.1$-$1000 $\mu m$ range, the middle panel the approximate range of the NIRspec instrument on \textit{JWST} and the right panel the corresponding range of the MIRI instrument.}
\label{fig:6508}
\end{figure*}

In Fig. \ref{fig:6508} we fit IRAS~07598+6508 with all four combinations of models without polar dust. We observe that in all cases there is evidence that the torus model by itself can not reproduce the (near-infrared) NIR spectrum. We attribute this to the presence of polar dust in this object, which is more visible because of the assumed face-on inclination. We can also see this from the reduced $\chi^2$ we provide in Table \ref{tab:6508nopolar}. The $\chi^{2}_{min, \nu}$ of the CYGNUS fit without polar dust is much higher compared to the fit with polar dust shown in Table A1. The important fitted parameters and extracted physical properties for this ULIRG are listed in Tables A1, A2, \ref{tab:extractedA} and \ref{tab:extractedB}. The SED fit plot with polar dust is shown in Fig. A2.

\begin{table}
	\centering
\caption{Reduced $\chi^2$ of fits for IRAS~07598+6508 with all combinations of models without using the polar dust component}
	\label{tab:6508nopolar}
 \begin{tabular}{ lcccccccc }
 \hline 
AGN model & $\chi^{2}_{min, \nu}$ \\ 
 \hline CYGNUS &   0.71  \\     
 \cite{fritz06}  &    0.89  \\   
 SKIRTOR &  0.66  \\     
 \cite{sieb15} &     2.82  \\   \hline 
 \end{tabular}
\end{table}

\subsection{Fit of the deeply obscured ULIRG IRAS~08572+3915}\label{3915}
\cite{efs14} proposed that the ULIRG IRAS~08572+3915 is the most luminous IR galaxy in the local ($z<0.2$) Universe. It is an example of an interacting ULIRG, which exhibits one of the deepest silicate absorption features observed in a galaxy. \cite{efs14} proposed that the deep silicate absorption feature is due to the fact that the AGN torus in this object is observed almost edge-on. The huge luminosity of this object, which approaches or exceeds $10^{13}L_\odot$, is due to the large anisotropy correction. 

To explore further the properties of this system, we fit the SED with all four combinations of AGN radiative transfer models. Two of the combinations assume a smooth torus geometry \citep{efstathiou95,fritz06} and two combinations assume a two-phase geometry \citep{sieb15,stal16}. 

\begin{figure*}
\centering
\includegraphics[width=.9\linewidth]{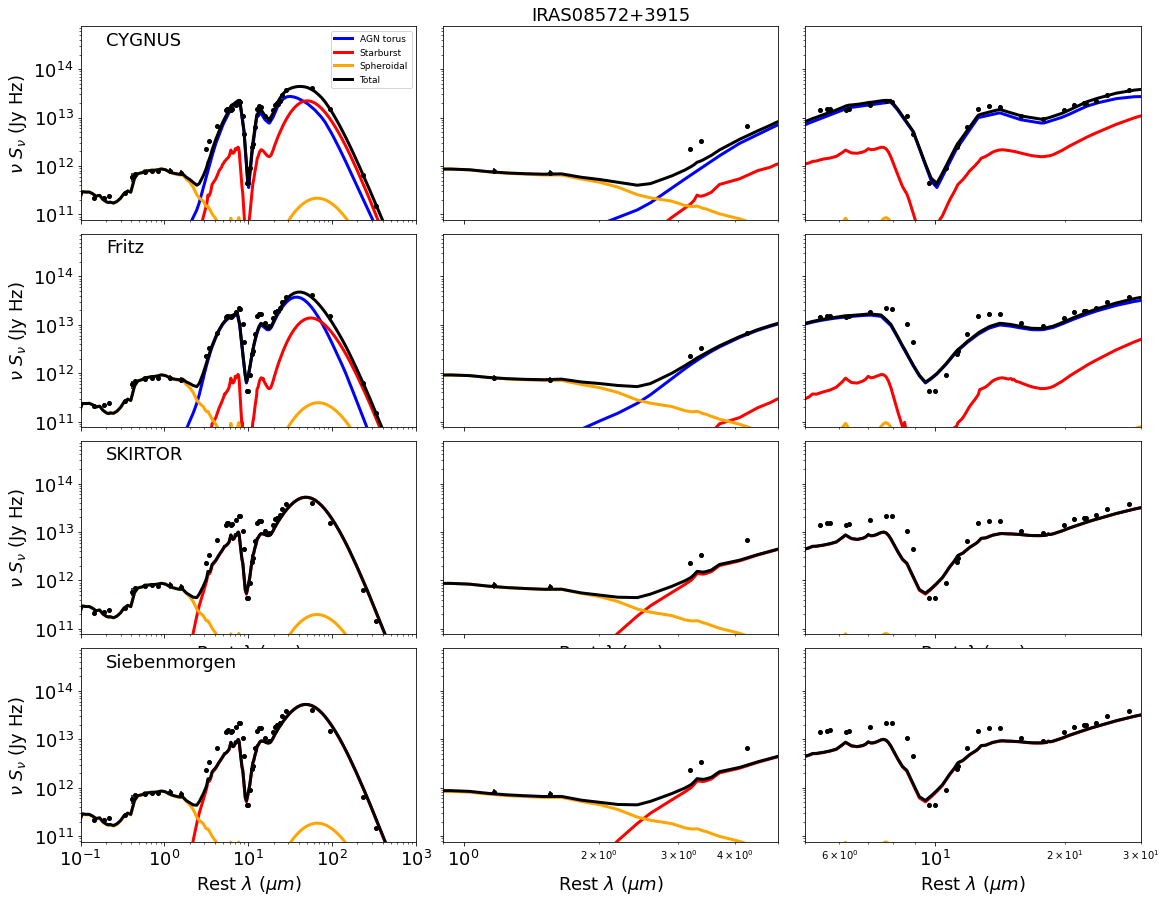}
\caption{Comparison SED fit plots of IRAS~08572+3915. The AGN torus, starburst, spheroidal host and total emission are plotted as shown in the legend. The top row shows fits with the CYGNUS combination of models. The second row shows fits with the CYGNUS AGN model replaced by the model of Fritz et al. (2006), the third row replaces the CYGNUS AGN model with the SKIRTOR model, while the bottom row replaces the CYGNUS AGN model with the model of Siebenmorgen et al. (2015). The left panel shows the fit over 0.1$-$1000 $\mu m$ range, the middle panel the approximate range of the NIRspec instrument on \textit{JWST} and the right panel the corresponding range of the MIRI instrument.}
\label{fig:3915}
\end{figure*}

It is interesting that the smooth models fit much better than the two-phase models. The last two in particular do not fit the deep silicate absorption feature of this galaxy and predict that the object is starburst-dominated. According to the fits of the smooth models, which perform much better, this is an AGN-dominated object. In Tables \ref{tab:3915par2}, A3 and A4 we list all of the important fitted parameters and extracted physical properties of the four combinations of models. The comparison SED fit plots are shown in Fig. \ref{fig:3915}. The CYGNUS combination of models gives a $\chi^{2}_{min, \nu}$ equal to 1.67, which is significantly lower than the $\chi^{2}_{min, \nu}$ given by the other three combinations and shows that the CYGNUS AGN model fits much better than the other models. The \cite{fritz06} AGN model gives $\chi^{2}_{min, \nu}=4.1$, the SKIRTOR AGN model gives $\chi^{2}_{min, \nu}=8.2$ and the AGN model of \cite{sieb15} gives $\chi^{2}_{min, \nu}=8.28$.

\begin{table*}
	\centering
\caption{Selected fitted parameters for IRAS~08572+3915}
	\label{tab:3915par2}
 \resizebox{\textwidth}{!}{\begin{tabular}{ lcccccccc }
 \hline 
AGN model  &   $t_{*} \ (10^7yr)$ & $\tau_{*} \ (10^7yr)$ &  & 
 &  &  $\theta_i \ (\degr)$ \\ 
 \hline CYGNUS &  &  & $r_2/r_1$ & $\tau_{uv}$ & $\theta_o \ (\degr)$ \\  
 $  $ &     $0.64^{0.28}_{-0.1}$  &    $1.79^{0.34}_{-0.25}$  &    $58.5^{9.8}_{-8.38}$  &    $612.99^{14.44}_{-21.13}$  &    $72.66^{1.05}_{-0.89}$  &    $79.04^{0.74}_{-0.65}$ \\ 
                                    &            & &  \\
 \cite{fritz06}  &  &  & $r_2/r_1$ & $\tau_{9.7\mu m}$ & $\theta_o \ (\degr)$ \\ 
 $  $ &    $2.37^{0.6}_{-0.6}$  &    $1.93^{0.17}_{-0.17}$  &    $117.4^{18.11}_{-18.11}$  &    $9.39^{0.03}_{-0.03}$  &    $54.48^{0.61}_{-0.61}$  &    $84.54^{0.01}_{-0.01}$ \\ 
                                    &            & &  \\
 SKIRTOR &  &  & $r_2/r_1$ & $\tau_{9.7\mu m}$ & $\theta_o \ (\degr)$ \\ 
 $  $ &    $0.51^{0.02}_{-0.01}$  &    $1.89^{0.53}_{-0.22}$  &    $23.96^{4.5}_{-4.72}$  &    $5.67^{2.13}_{-0.7}$  &    $47.31^{16.22}_{-11.34}$  &    $60.72^{13.35}_{-13.6}$ \\  
                                    &            & &  \\
 \cite{sieb15} &  &  & $A_c$ & $V_c$ & $A_d$ \\ 
 $  $ &    $0.51^{0.02}_{-0.01}$  &    $2.05^{0.74}_{-0.25}$  &    $8.36^{14.42}_{-6.23}$  &    $57.98^{47.19}_{-21.28}$  &    $211.06^{126.38}_{-58.5}$  &    $86.88^{6.87}_{-2.63}$ \\   \hline 
 \end{tabular}}
\end{table*}

\subsection{Fit of a LIRG associated with a spiral galaxy}
To explore the option of fitting with a disc as a host galaxy component, we fitted the LIRG VV~340a, which belongs to a pair of spiral galaxies known as Arp~302, with the northern galaxy, VV~340a, being viewed edge-on and the southern galaxy, VV~340b, viewed face-on. This pair of galaxies was discussed extensively by \cite{armus09}. We fitted this object twice, using the two different host galaxy models that we have available, the spheroidal and the disc model.

\begin{figure*}
\centering
\includegraphics[width=.9\linewidth]{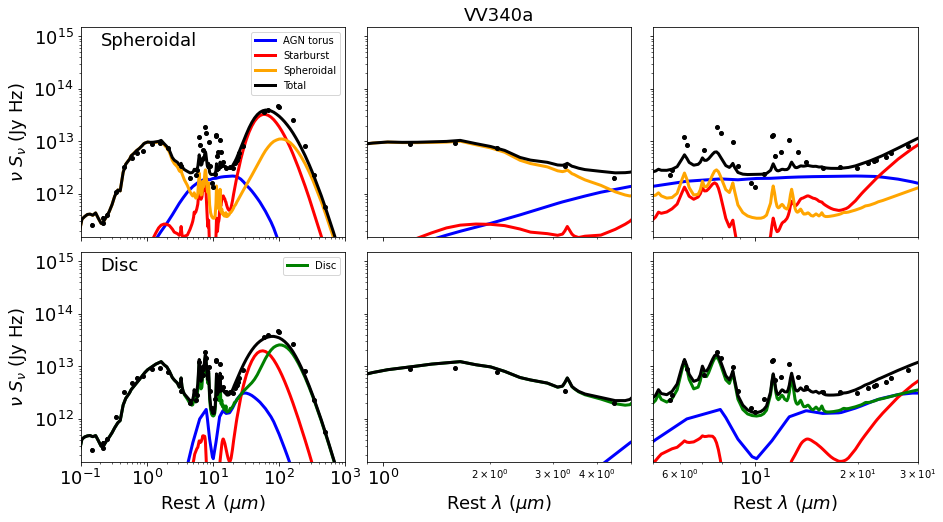}
\caption{Comparison SED fit plots of VV~340a, which is a LIRG associated with a spiral galaxy. The AGN torus, starburst, host and total emission are plotted
as shown in the legend. The top row shows the fit with the spheroidal host galaxy model, while the second row replaces the spheroidal host galaxy model with the disc host galaxy model. The left panel shows the fit over 0.1$-$1000 $\mu m$ range, the middle panel the approximate range of the NIRspec instrument on \textit{JWST} and the right panel the corresponding range of the MIRI instrument.}
\label{fig:vv340a}
\end{figure*}

We observe that the disc model appears to fit better not just the photometry but also the Spitzer spectroscopy. The difference in the $\chi^{2}_{min, \nu}$ of the two different fits is also significant, with the disc model providing the minimum value of 2.84, while the spheroidal model provides a $\chi^{2}_{min, \nu}$ of 5.65, a factor of two higher. The best fits with the two different host geometries are listed in Tables \ref{tab:vv340apar2}, A5 and A6 and plotted in Fig. \ref{fig:vv340a}. The predicted disc morphology is consistent with the nature of this galaxy. It is very interesting that the fit of VV~340a predicts an almost edge-on view and this agrees with the image of the galaxy. This example is an indication that SMART may be a useful tool for galaxy morphological classification, especially in cases where an image of the galaxy is not available.

\cite{jiang24} presented XMM-Newton data of VV~340a and VV~340b from which they find evidence of a heavily obscured Seyfert 2-like AGN in VV~340a. This is in agreement with our fit, which predicts the presence of an obscured AGN with an anisotropy-corrected AGN luminosity of $0.25^{0.06}_{-0.04} \times 10^{12} L_\odot$. Our model, however, predicts a much higher SFR compared with the estimate of \cite{jiang24} of 1$-$2 $M_\odot$/yr for this galaxy, which is based entirely on X-ray data.

\begin{table*}
	\centering
\caption{Selected fitted parameters for VV~340a}
	\label{tab:vv340apar2}
 \resizebox{\textwidth}{!}{\begin{tabular}{ lccccccc }
 \hline 
Host galaxy model & $t_{*} \ (10^7yr)$ & $\tau_{*} \ (10^7yr)$ & $r_2/r_1$ & $\tau_{uv}$ & $\theta_o \ (\degr)$ & $\theta_i \ (\degr)$ \\ \hline 
 Spheroidal  &  $3.38^{0.02}_{-0.16}$  &    $1.46^{0.26}_{-0.22}$  &    $78.02^{4.58}_{-6.31}$  &    $760.03^{342.98}_{-182.2}$  &    $56.9^{3.32}_{-3.09}$  &    $55.81^{3.95}_{-4.56}$ \\ 
                                     &            & &  \\
 Disc &   $1.09^{0.1}_{-0.07}$  &    $2.02^{0.05}_{-0.09}$  &    $89.72^{5.24}_{-5.15}$  &    $863.0^{117.51}_{-51.98}$  &    $66.53^{1.41}_{-1.56}$  &    $74.07^{0.53}_{-0.66}$ \\   \hline 
 \end{tabular}}
\end{table*}

\subsection{Fit of a hyperluminous obscured quasar at $z\sim4.3$} \label{help}
The HLIRG HELP\_J100156.75+022344.7 at a photometric redshift of $z\sim4.3$ was discovered in the Cosmological Evolution Survey (COSMOS) field, one of the fields studied by HELP. This is one of only a few $z>4$ hyperluminous obscured quasars discovered to date and, according to the CYGNUS model presented by \cite{efs21}, it is AGN-dominated with an AGN fraction of 82 per cent. \cite{efs21} also fitted HELP\_J100156.75+022344.7 with CIGALE and arrived at a similar conclusion. HELP\_J100156.75+022344.7 is predicted to have an SFR of $1051^{109}_{-268}$ $M_\odot yr^{-1}$ according to CYGNUS and $991^{223}_{-223}$ $M_\odot yr^{-1}$ according to CIGALE.

In this paper we fit HELP\_J100156.75+022344.7 with the four different combinations of models, in order to quantify the uncertainty in the AGN fraction and SFR. As noted by \cite{efs21}, in the SUPRIME N921 filter we see an excess in the photometry that deviates significantly from the fit. This could be due to an emission line or a problem in the data reduction. As in the models we use in our method emission lines are not included, we opted to omit this data point from the fit.

\begin{figure}
\centering
\includegraphics[width=.9\linewidth]{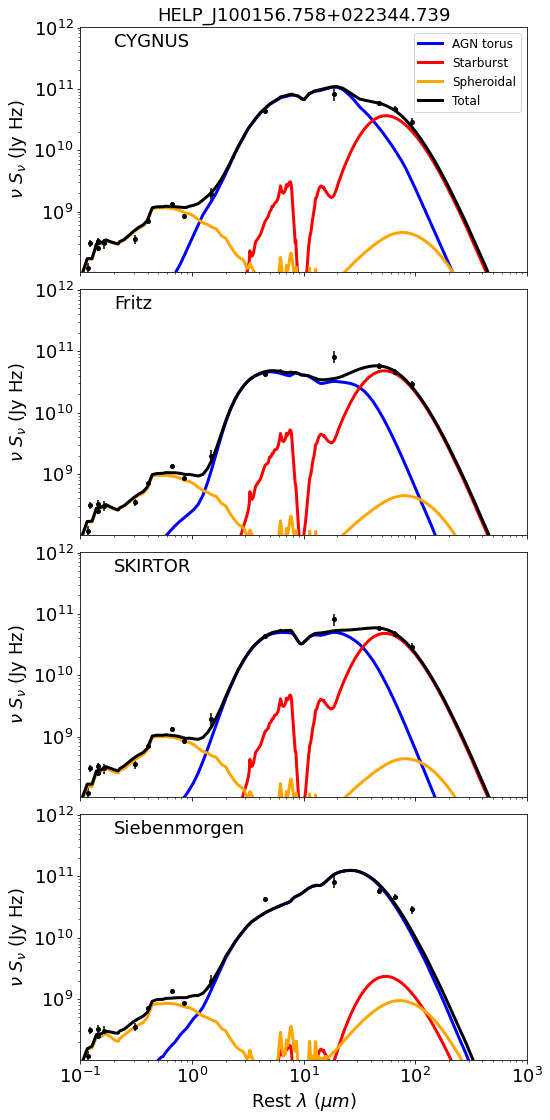}
\caption{Comparison SED fit plots of the hyperluminous galaxy HELP\_J100156.75+022344.7 at a photometric redshift of $z\sim4.3$. The AGN
torus, starburst, spheroidal host and total emission are plotted as shown in the
legend. The top row shows fits with the CYGNUS combination of models. The second row shows fits with the CYGNUS AGN model replaced by the model of Fritz et al. (2006), the third row replaces the CYGNUS AGN model with the SKIRTOR model, while the bottom row replaces the CYGNUS AGN model with the model of Siebenmorgen et al (2015).}
\label{fig:help}
\end{figure}

For all the fits we fixed $\tau_{*} = 2 \times 10^7yr$ and two other parameters. For the fits with the CYGNUS and \cite{fritz06} AGN models we additionally fixed $\theta_o = 45 \degr$ and $r_2/r_1 = 60$. For the fit with the AGN model SKIRTOR we fixed $\theta_o = 45 \degr$ and $r_2/r_1 = 20$, whereas for the fit with the \cite{sieb15} AGN model we fixed $A_c = 3$ and $V_c = 40$. The best fits with all combinations of models are listed in Tables \ref{tab:helppar2}, A7 and A8 and plotted in Fig. \ref{fig:help}. The CYGNUS, \cite{fritz06} and SKIRTOR models predict a similar solution for the AGN fraction and SFR, although there are significant differences in the $\chi^{2}_{min, \nu}$, with the CYGNUS model giving the minimum value of 6.09. The SKIRTOR model gives $\chi^{2}_{min, \nu}=6.69$ and the \cite{fritz06} model gives $\chi^{2}_{min, \nu}=8.67$. The model of \cite{sieb15} gives the worst fit with $\chi^{2}_{min, \nu}=12.03$ and predicts a significantly lower SFR compared with the other models.

\begin{table*}
	\centering
\caption{Selected fitted parameters for HELP\_J100156.75+022344.7}
	\label{tab:helppar2}
 \begin{tabular}{ lcccccccc }
 \hline 
AGN model & $\tau_v$ &   $t_{*} \ (10^7yr)$ & & $\theta_i \ (\degr)$ \\ 
 \hline CYGNUS &  &  & $\tau_{uv}$ \\ 
 $  $ &    $125.81^{34.15}_{-17.31}$ &    $2.02^{0.19}_{-0.71}$  &    $367.97^{207.86}_{-50.55}$  &    $55.18^{0.48}_{-1.8}$ \\  
                                     &            & &  \\
 \cite{fritz06}  &  &  & $\tau_{9.7\mu m}$ \\ 
 $  $ &    $107.94^{33.15}_{-30.26}$ &    $1.6^{0.45}_{-0.44}$  &    $2.2^{1.6}_{-0.64}$  &    $79.08^{4.53}_{-4.11}$ \\  
                                     &            & &  \\
 SKIRTOR &  &  & $\tau_{9.7\mu m}$ \\ 
 $  $ &    $114.31^{92.24}_{-41.79}$ &     $1.82^{1.05}_{-0.79}$  &    $6.01^{1.4}_{-2.12}$  &    $64.79^{5.57}_{-12.75}$ \\   
                                     &            & &  \\
 \cite{sieb15} &  &  & $A_d$ \\  
 $  $ &    $133.18^{54.28}_{-45.45}$  &    $1.53^{0.82}_{-0.76}$  &    $136.07^{37.17}_{-58.89}$  &    $60.1^{5.86}_{-6.16}$ \\   \hline 
 \end{tabular}
\end{table*}

\section{Conclusions}\label{conclusion}

\begin{enumerate}

\item In this paper we present SMART, a new fast code of fitting SEDs exclusively with radiative transfer models. The method employs the MCMC code \textit{emcee} and uses pre-computed libraries for starbursts, AGN tori and host galaxies, which can be modelled either with a spheroidal or disc geometry. The code takes comparable time to energy balance codes like CIGALE and MAGPHYS. The method is flexible in that it can fit SEDs which include MIR spectrophotometry, but also galaxies where more limited photometry is available.

\item As SMART can fit an SED with four different AGN torus models, it allows us to constrain the properties of the obscuring torus in AGN and quantify the uncertainties in AGN fraction and SFR. We consider four different AGN torus models: the smooth tapered discs of \cite{efstathiou95}, the smooth flared discs of \cite{fritz06}, the two-phase flared discs of \cite{sta12,stal16} and the two phase model of \cite{sieb15}. The first three models assume a normal interstellar mixture, whereas the model of \cite{sieb15} assumes fluffy grains, which are more efficient at emitting at submillimetre wavelengths.

\item We test the method with the HERUS sample of ULIRGs and compare our results with those obtained with the SATMC code by \cite{efs22} and by \cite{pasp21} using CIGALE. We find that the results obtained with SMART agree very well with those obtained with SATMC, except for the AGN fraction, which shows considerable scatter. We attribute this to the strong dependence of this quantity on the inclination. The SFRs predicted by CIGALE generally agree with those with SMART, but the AGN fraction is significantly lower, partly because CIGALE does not apply any anisotropy corrections to the AGN luminosity.

\item SMART is developed in PYTHON and is available at \url{https://github.com/ch-var/SMART.git}.

\end{enumerate}

We also briefly discuss some ideas for future development of SMART:

\begin{enumerate}

\item The main limitation of SMART is that it currently assumes constant metallicity and a parametric SFH and in particular a delayed exponential for the host galaxy and an exponential for the starburst. Future developments of the code will allow treatment of non-parametric SFH and metallicity histories.

\item Another limitation of SMART is that it does not model nebular lines. A future development could be to add nebular emission in the libraries used by SMART. To do this, we would need to increase the spectral resolution of the models, which is currently 223 points, to cover the wavelength range from the UV to the millimetre.

\item SMART currently fits the UV to millimetre SEDs. Therefore, another limitation is that it does not model X-ray emission. This could be an additional extension of SMART. We could follow the same methodology adopted in other codes, such as X-CIGALE \citep{yang20}.

\end{enumerate}

We conclude that SMART promises to be very useful for understanding galaxies and AGN at any redshift in the \textit{JWST} era, as well as galaxy formation and evolution more generally.

\section*{Data Availability Statement}
The data underlying this article are available in the article, in public databases Cornell Atlas of Spitzer/Infrared Spectrograph Sources (CASSIS) or are publicly available in the literature.

\section*{Acknowledgements}

We would like to thank the anonymous referee for useful comments and suggestions. The authors acknowledge support from the projects CYGNUS (contract number 4000126896) and CYGNUS+ (contract number 4000139319) funded by the \textcolor{cyan}{European Space Agency}. CV acknowledges a Ph.D. scholarship from European University Cyprus.

%%%%%%%%%%%%%%%%%%%%%%%%%%%%%%%%%%%%%%%%%%%%%%%%%%

%%%%%%%%%%%%%%%%% APPENDICES %%%%%%%%%%%%%%%%%%%%%

%%%%%%%%%%%%%%%%%%%%%%%%%%%%%%%%%%%%%%%%%%%%%%%%%%
\section*{Appendix A: Plots and other SED fitting results discussed in the paper} 

In this Appendix we give the plots of all of the SED fits with the CYGNUS models, as well as their residuals, for the HERUS sample. We also give two tables with the best fit parameters for the fits with the CYGNUS models for the HERUS sample. In addition, we provide tables with extracted physical quantities of the fits of IRAS~08572+3915, VV~340a and HELP\_J100156.75+022344.7. 

\begin{figure*}
	\begin{center}
      {\includegraphics[width=80mm]{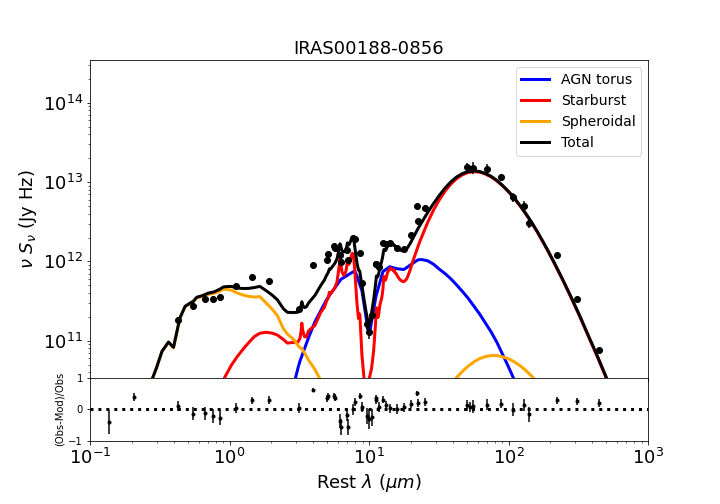}}    
      {\includegraphics[width=80mm]{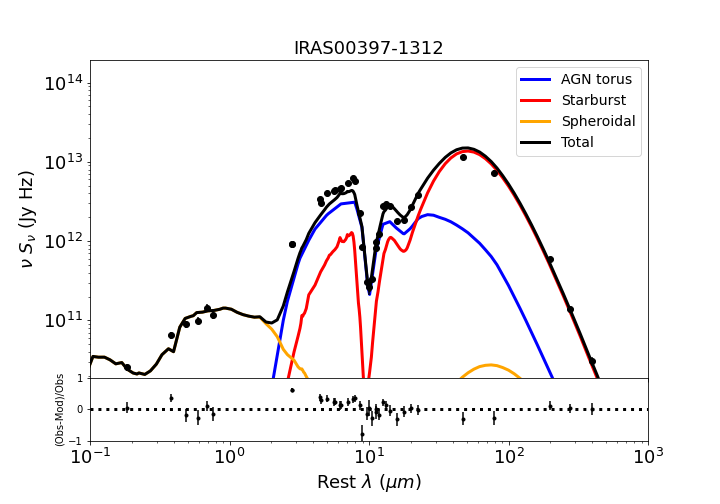}}    
      {\includegraphics[width=80mm]{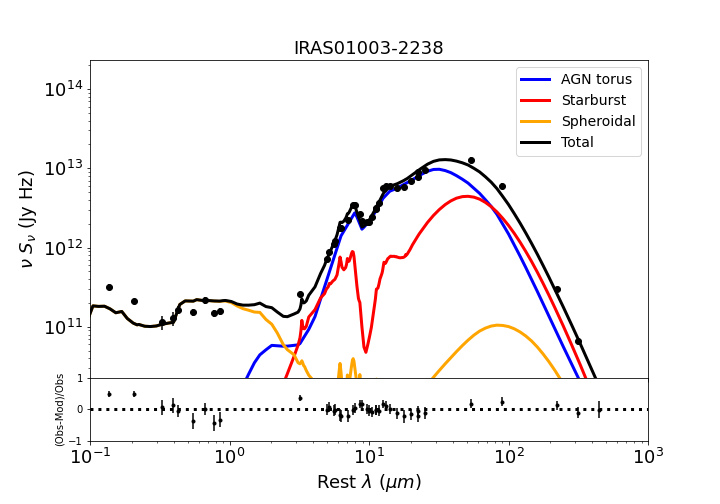}} 
      {\includegraphics[width=80mm]{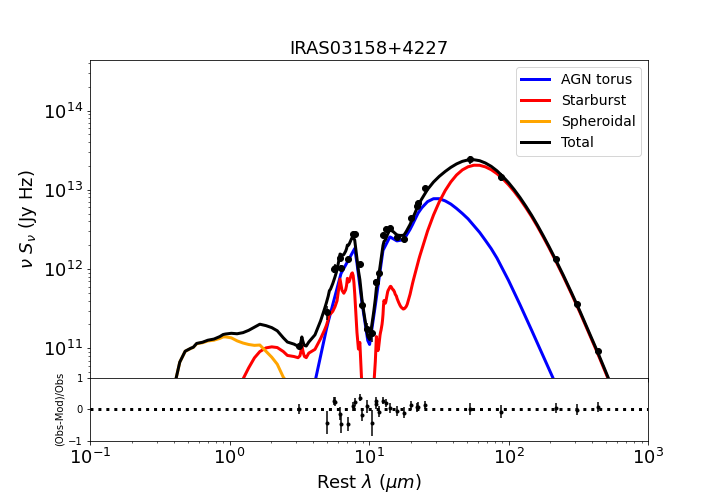}} 
      {\includegraphics[width=80mm]{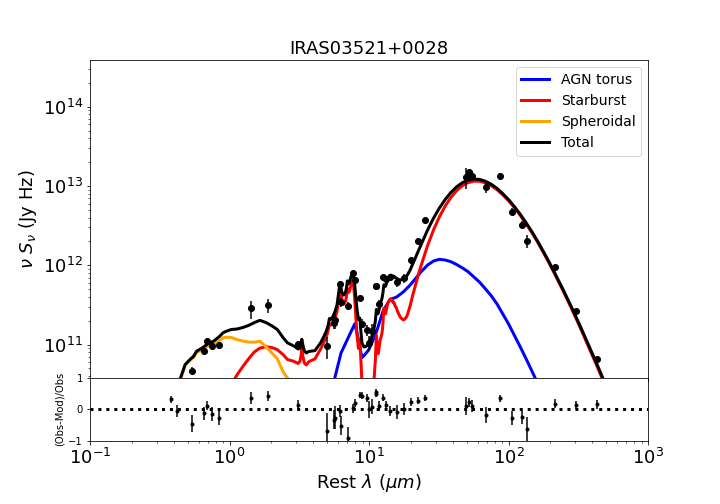}}      
      {\includegraphics[width=80mm]{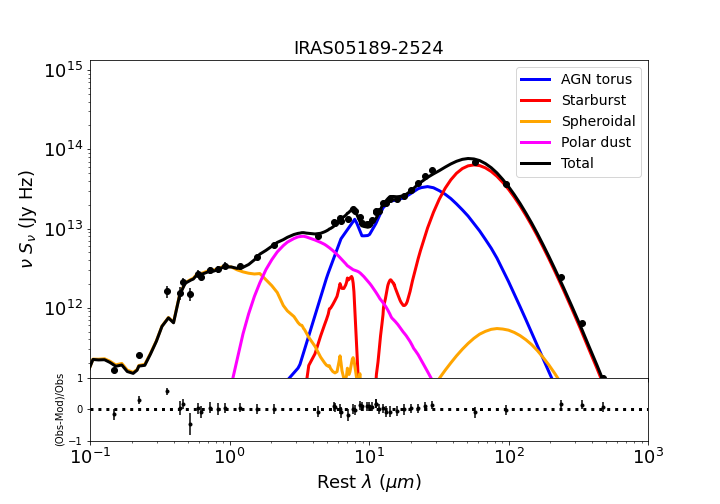}} 
      {\includegraphics[width=80mm]{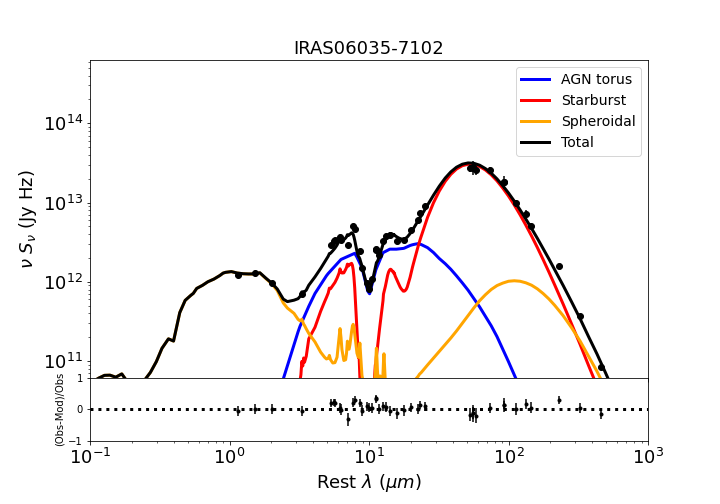}} 
      {\includegraphics[width=80mm]{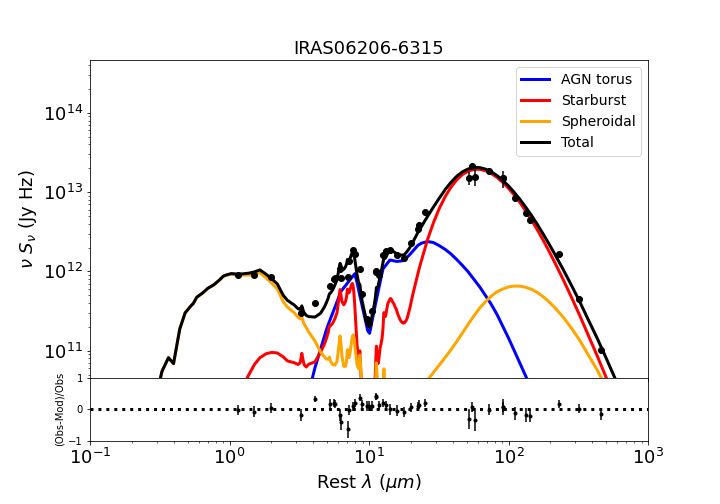}} 
      \\
\begin{flushleft}
\textbf{Figure A1.} SED fit plots of the first 8 objects from the list of the HERUS sample, using the CYGNUS models. The AGN torus, starburst, spheroidal host, polar dust and total emission are plotted as shown in the legend.
\end{flushleft}
     \label{fig:HERUS-resultsA}
\end{center}
\end{figure*}

\begin{figure*}
	\begin{center}
      {\includegraphics[width=80mm]{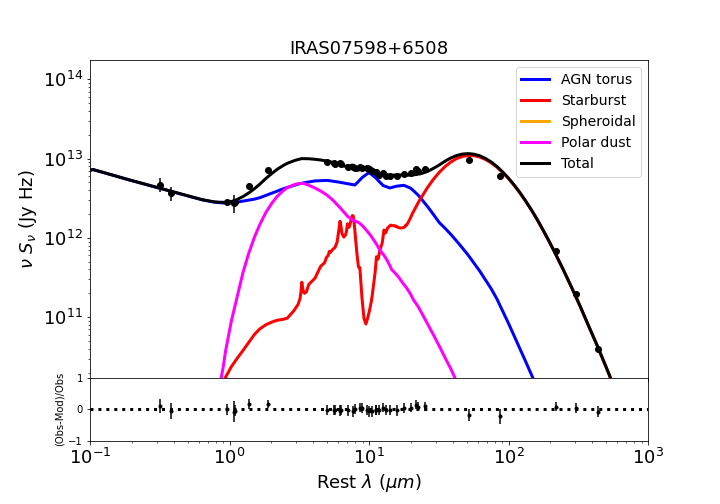}} 
      {\includegraphics[width=80mm]{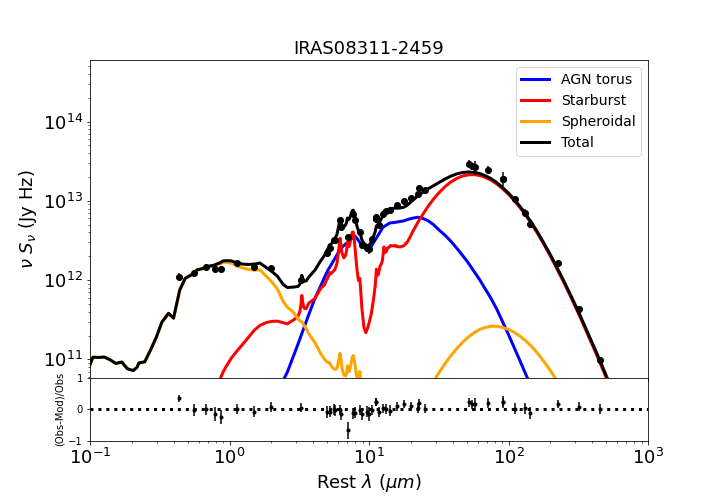}} 
      {\includegraphics[width=80mm]{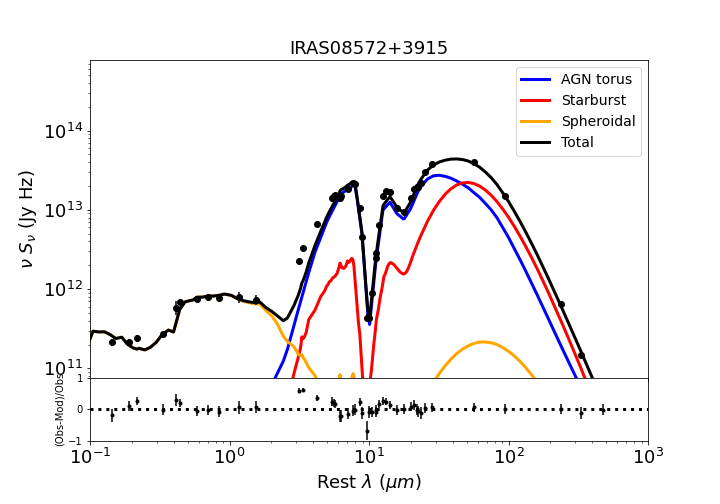}} 
      {\includegraphics[width=80mm]{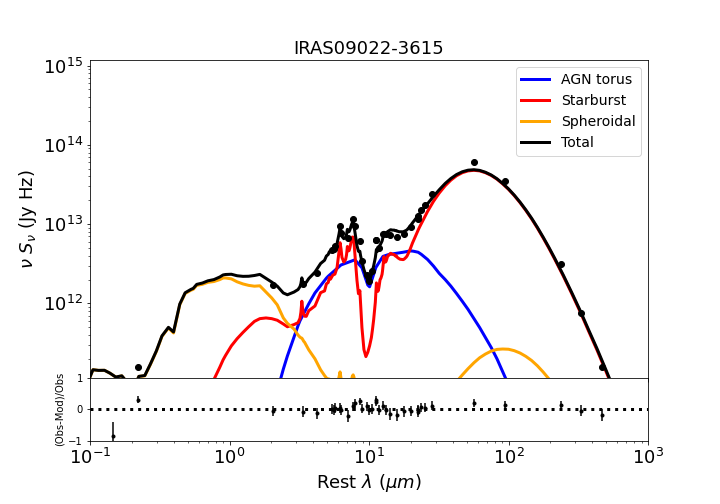}} 
      {\includegraphics[width=80mm]{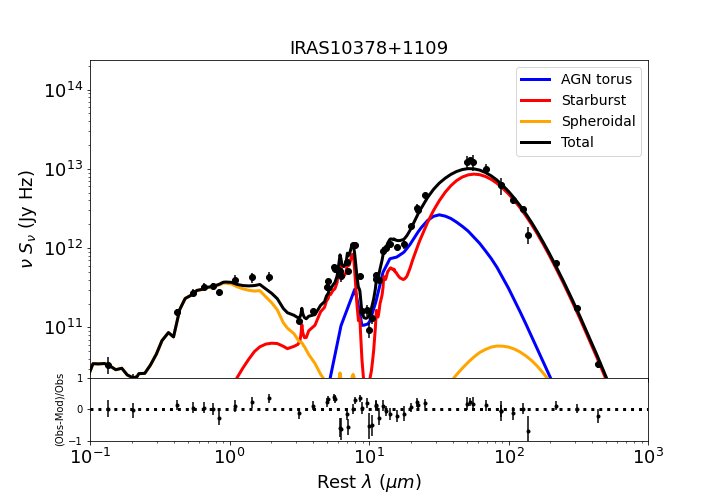}}
      {\includegraphics[width=80mm]{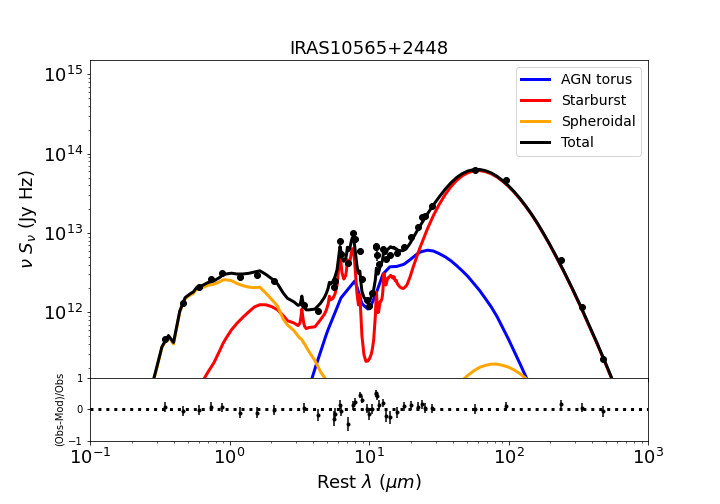}} 
      {\includegraphics[width=80mm]{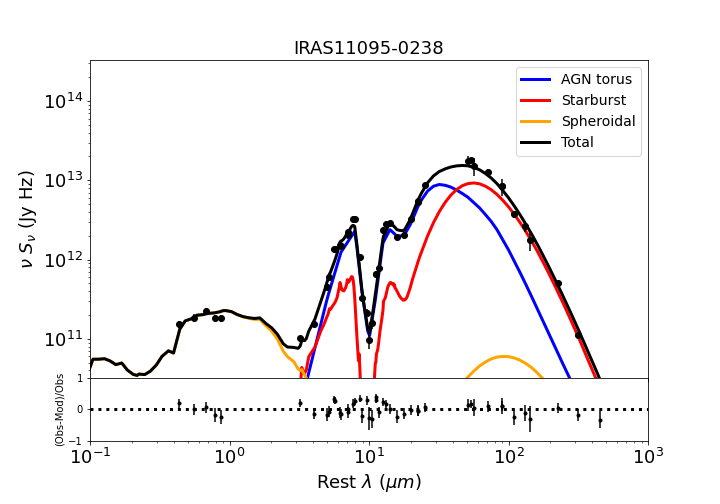}} 
      {\includegraphics[width=80mm]{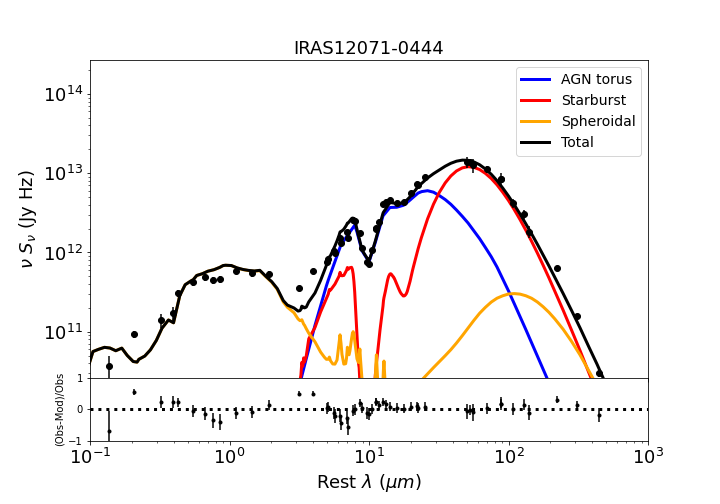}} 
      \\
\begin{flushleft}
\textbf{Figure A2.} SED fit plots of the second 8 objects from the list of the HERUS sample, using the CYGNUS models. The AGN torus, starburst, spheroidal host, polar dust and total emission are plotted as shown in the legend.
\end{flushleft}
     \label{fig:HERUS-resultsB}
\end{center}
\end{figure*}

\begin{figure*}
	\begin{center}
      {\includegraphics[width=80mm]{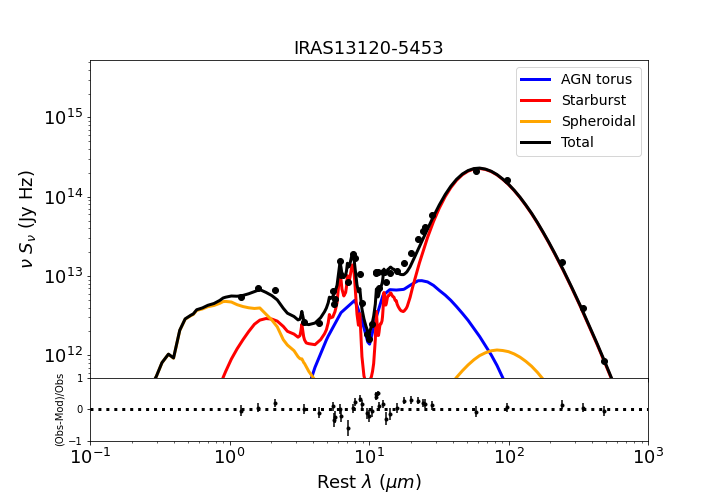}} 
      {\includegraphics[width=80mm]{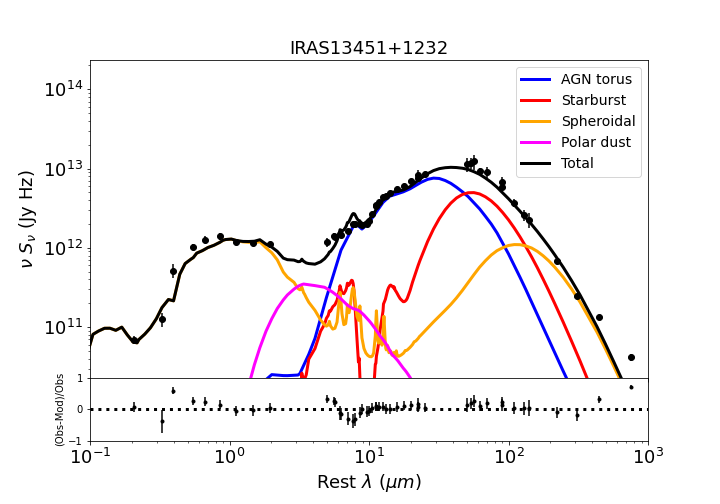}}      
      {\includegraphics[width=80mm]{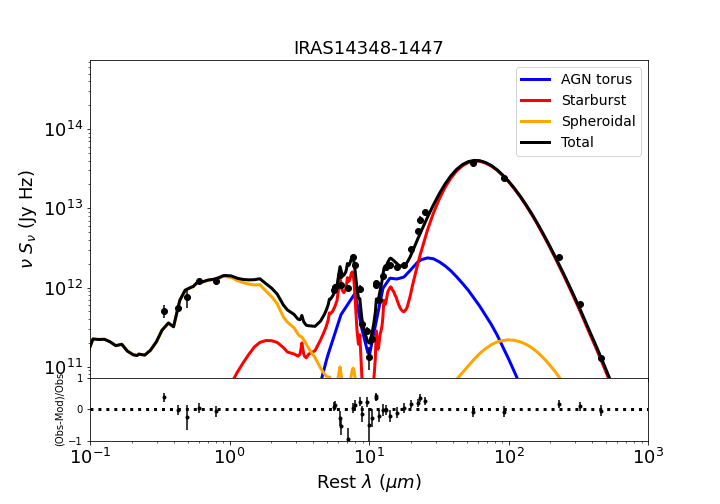}} 
      {\includegraphics[width=80mm]{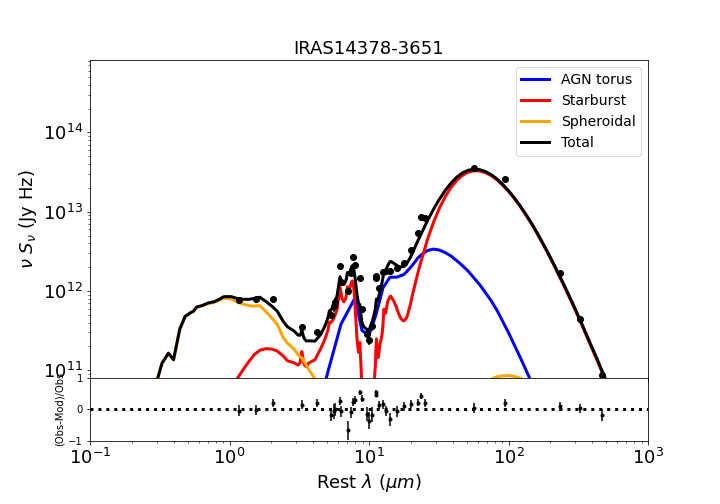}}     
      {\includegraphics[width=80mm]{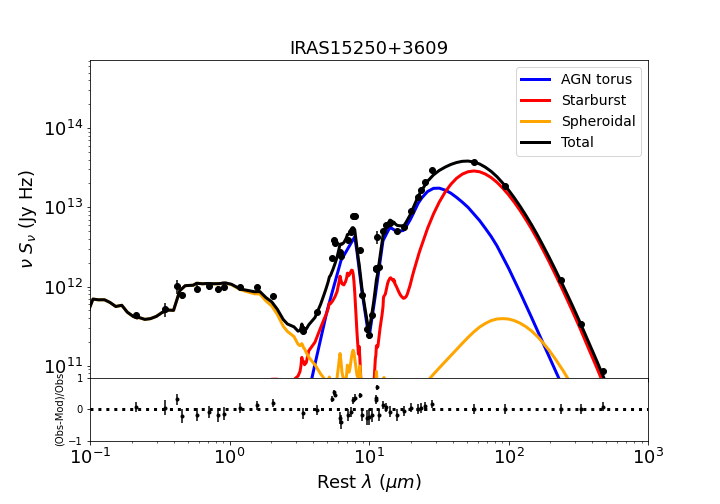}} 
      {\includegraphics[width=80mm]{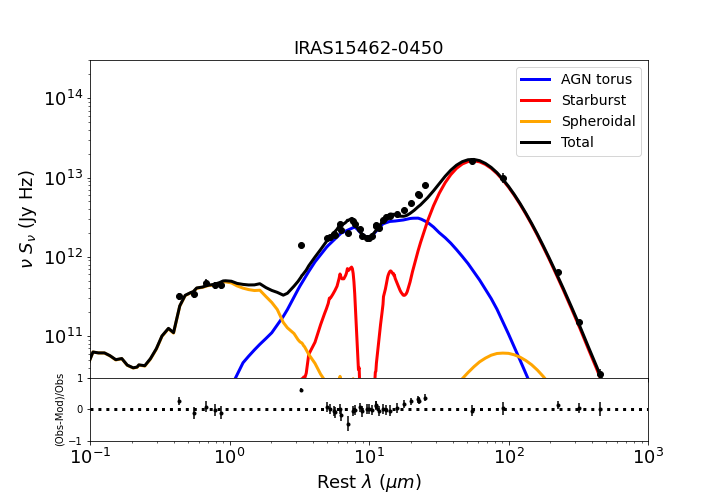}}     
      {\includegraphics[width=80mm]{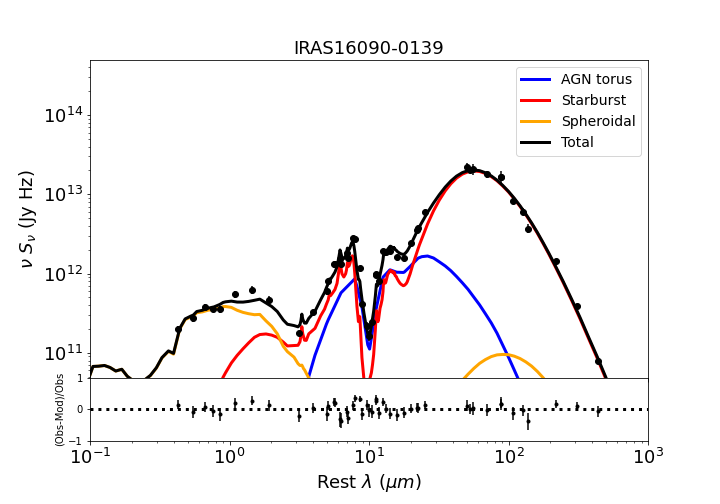}} 
      {\includegraphics[width=80mm]{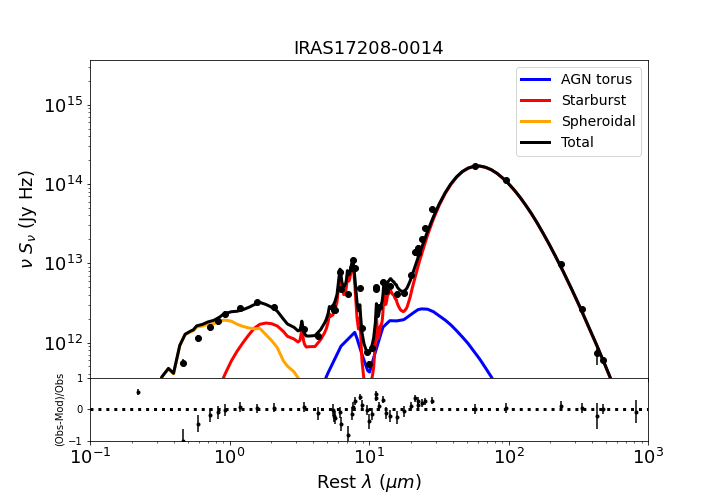}} 
      \\ 
\begin{flushleft}      
\textbf{Figure A3.} SED fit plots of the third 8 objects from the list of the HERUS sample. The AGN torus, starburst, spheroidal host,
polar dust and total emission are plotted as shown in the legend.
\end{flushleft}
     \label{fig:HERUS-resultsC}
\end{center}
\end{figure*}
 
\begin{figure*}
	\begin{center}
      {\includegraphics[width=80mm]{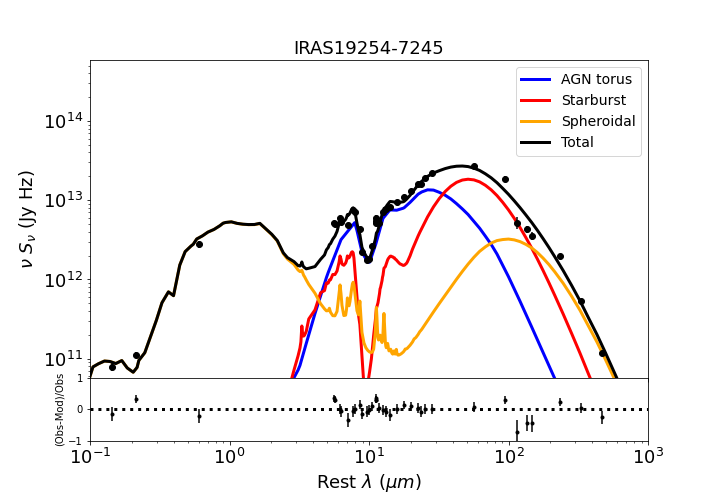}} 
      {\includegraphics[width=80mm]{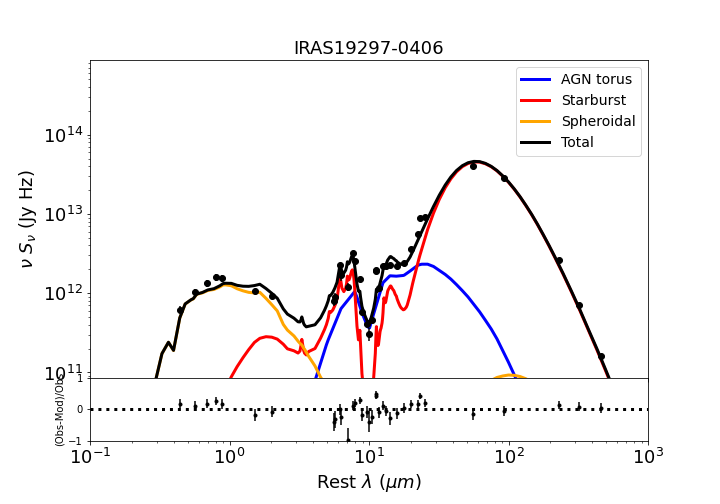}}     
      {\includegraphics[width=80mm]{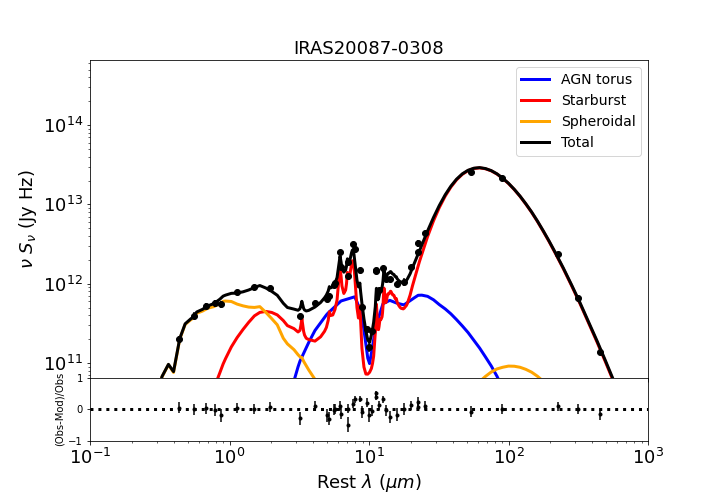}} 
      {\includegraphics[width=80mm]{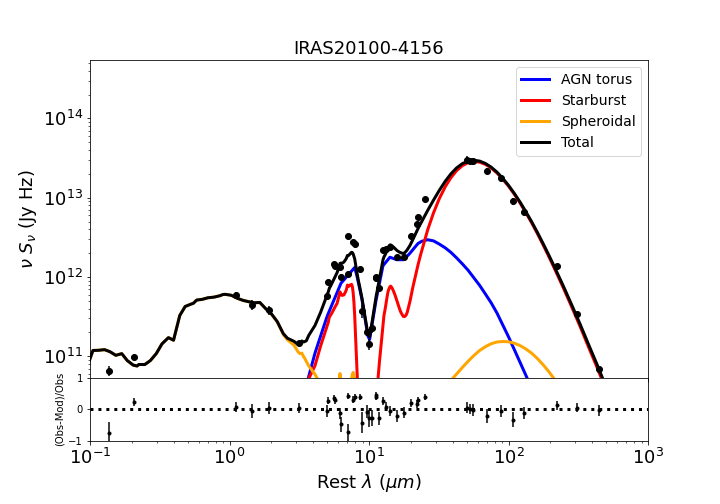}}      
      {\includegraphics[width=80mm]{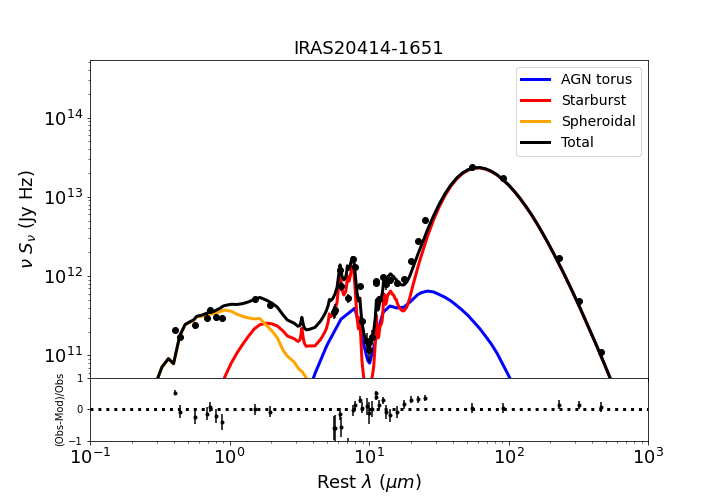}} 
      {\includegraphics[width=80mm]{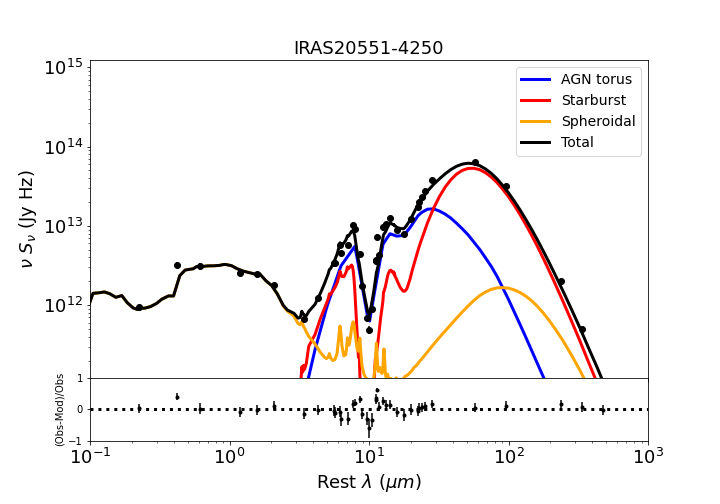}} 
      {\includegraphics[width=80mm]{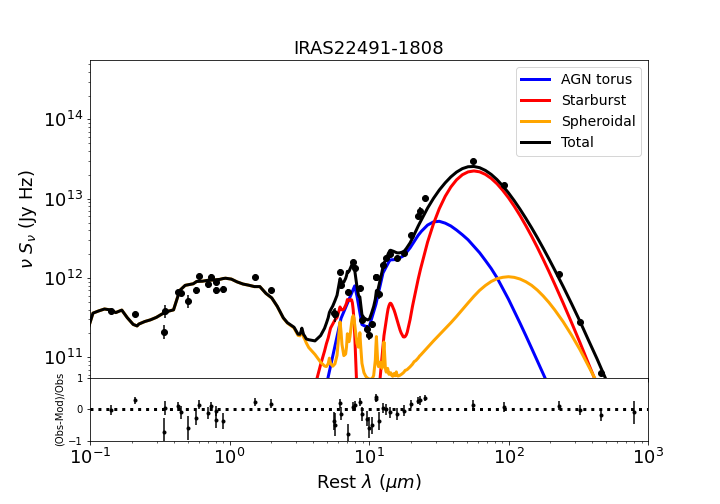}} 
      {\includegraphics[width=80mm]{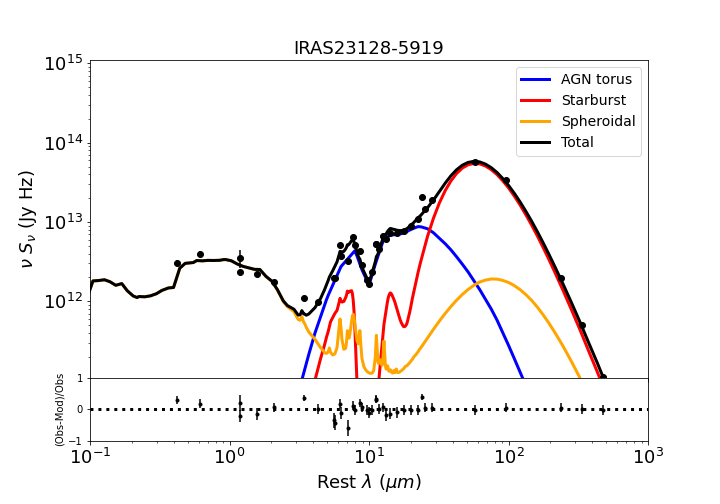}}  
      \\
\begin{flushleft}      
\textbf{Figure A4.} SED fit plots of the fourth 8 objects from the list of the HERUS sample. The AGN torus, starburst, spheroidal host and total emission are plotted as shown in the legend.
\end{flushleft}
     \label{fig:HERUS-resultsD}
\end{center}
\end{figure*}

\begin{figure*}
	\begin{center}
      {\includegraphics[width=80mm]{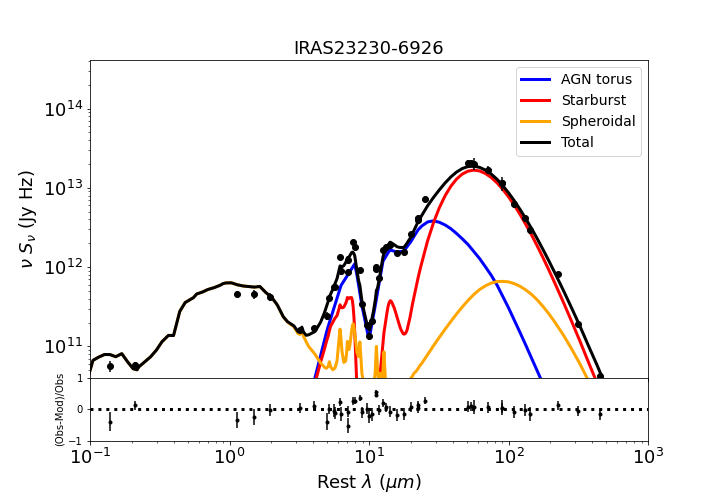}} 
      {\includegraphics[width=80mm]{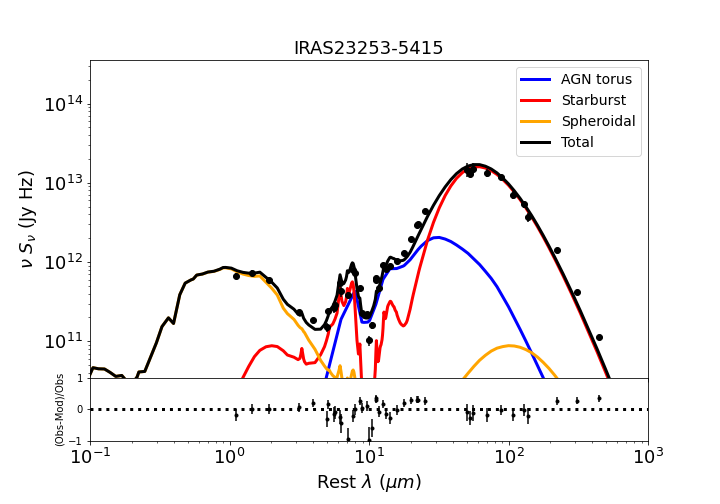}}       
      {\includegraphics[width=80mm]{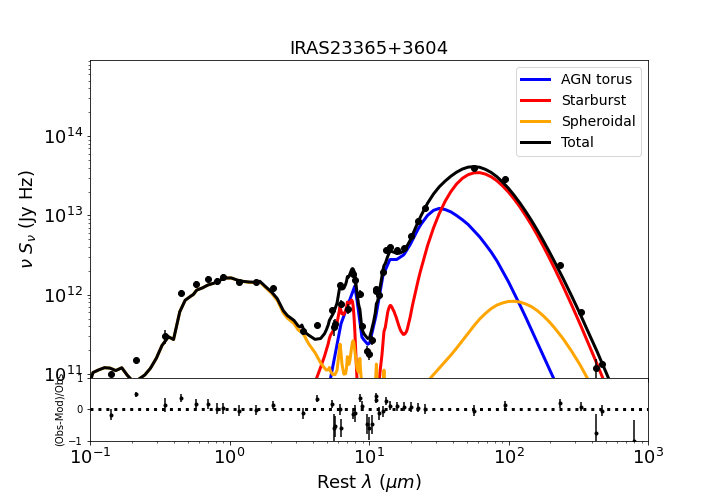}} 
      {\includegraphics[width=80mm]{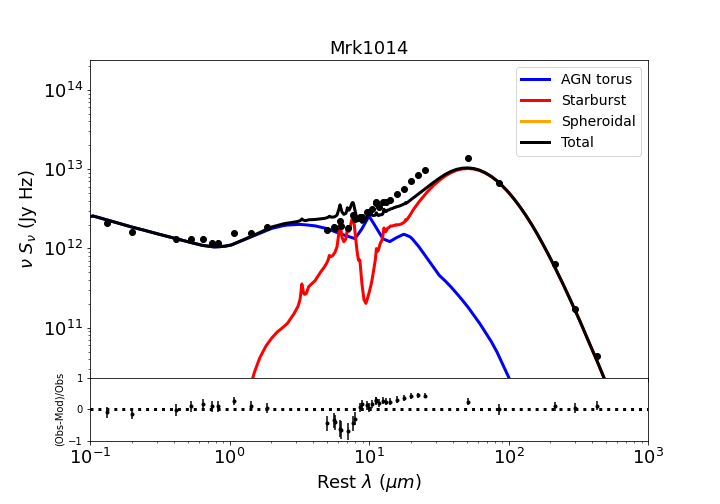}}          
      {\includegraphics[width=80mm]{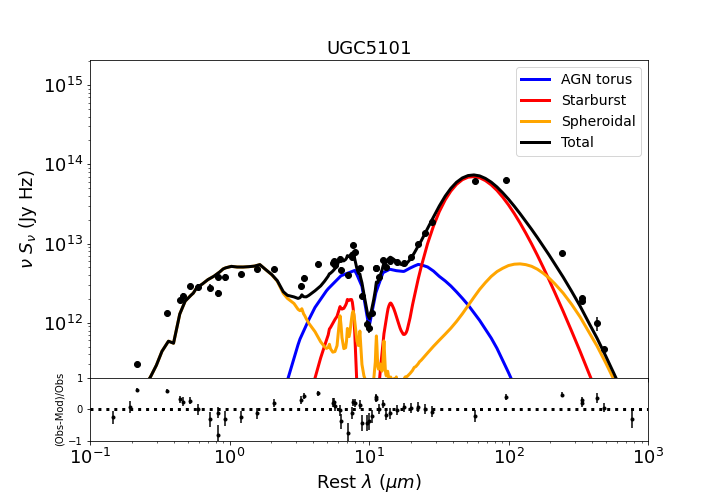}} 
      {\includegraphics[width=80mm]{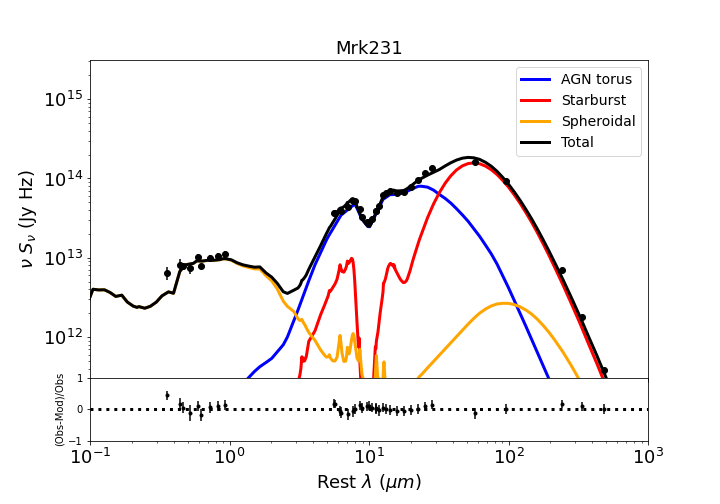}} 
      {\includegraphics[width=80mm]{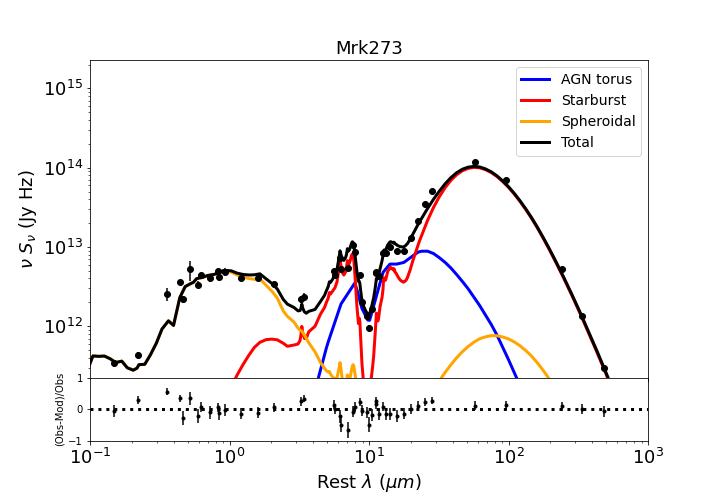}} 
      {\includegraphics[width=80mm]{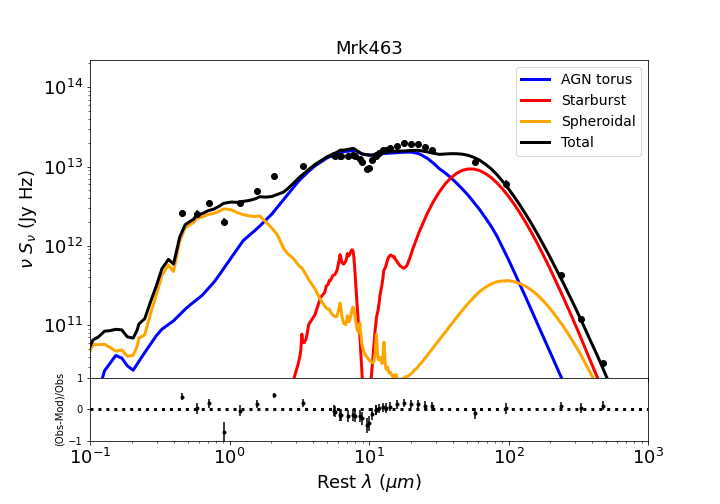}} 
      \\
\begin{flushleft}      
\textbf{Figure A5.} SED fit plots of the fifth 8 objects from the list of the HERUS sample. The AGN torus, starburst, spheroidal host and total emission are plotted as shown in the legend.
\end{flushleft}
     \label{fig:HERUS-resultsE}
\end{center}
\end{figure*}

\begin{figure*}
	\begin{center}
      {\includegraphics[width=80mm]{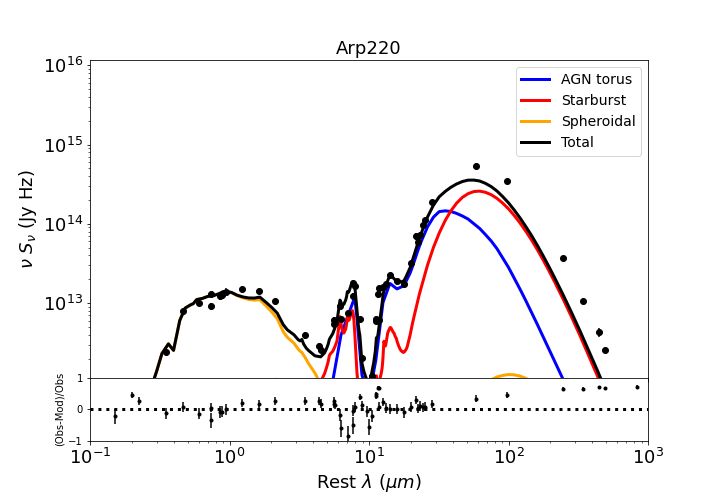}} 
      {\includegraphics[width=80mm]{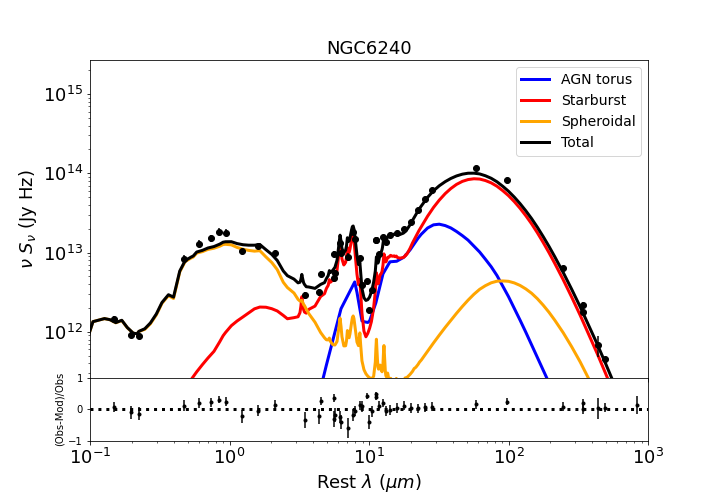}} 
      \\
\begin{flushleft}      
\textbf{Figure A6.} SED fit plots of the last 2 objects from the list of the HERUS sample. The AGN torus, starburst, spheroidal host and total emission are plotted as shown in the legend.
\end{flushleft}
     \label{fig:HERUS-resultsF}
\end{center}
\end{figure*}

\begin{table*}
	\centering
\begin{flushleft}
\textbf{Table A1.} Reduced $\chi^2$ and selected fitted parameters for the galaxies in the HERUS sample
\end{flushleft}
	\label{tab:fittedA}
 \resizebox{\textwidth}{!}{\begin{tabular}{ lcccccccc }
 \hline 
Name & $z$ & $\chi^{2}_{min, \nu}$ & $\tau_{v}^s$ & $\psi^s$ & $\tau^s \ (10^7yr)$ & $\tau_v$ \\ 
 \hline IRAS00188-0856 & 0.12840001 &    2.66  &    $1.23^{11.34}_{-0.61}$  &    $5.59^{0.88}_{-0.92}$  &    $147.11^{60.13}_{-9.47}$  &    $128.66^{9.16}_{-17.77}$ \\  IRAS00397-1312 & 0.2617 &    2.78  &    $0.44^{0.54}_{-0.24}$  &    $8.92^{6.17}_{-2.37}$  &    $326.65^{50.05}_{-65.95}$  &    $105.47^{5.39}_{-3.12}$ \\  IRAS01003-2238 & 0.1178 &    1.3  &    $0.77^{0.17}_{-0.36}$  &    $2.56^{0.4}_{-0.4}$  &    $778.94^{6.75}_{-29.1}$  &    $65.58^{4.93}_{-7.73}$ \\  IRAS03158+4227 & 0.1344 &    1.64  &    $0.44^{0.67}_{-0.2}$  &    $3.07^{2.46}_{-1.15}$  &    $178.15^{208.11}_{-127.35}$  &    $186.57^{12.27}_{-14.86}$ \\  IRAS03521+0028 & 0.1519 &    3.16  &    $4.42^{4.06}_{-1.95}$  &    $2.28^{0.9}_{-0.45}$  &    $25.37^{3.22}_{-4.22}$  &    $179.99^{14.26}_{-15.5}$ \\  IRAS05189-2524 & 0.04256 &    0.89  &    $0.91^{0.45}_{-0.23}$  &    $6.18^{1.3}_{-1.44}$  &    $166.31^{3.02}_{-5.19}$  &    $176.84^{12.32}_{-17.3}$ \\  IRAS06035-7102 & 0.07947 &    0.87  &    $9.24^{1.4}_{-5.56}$  &    $3.7^{0.86}_{-0.28}$  &    $205.86^{42.66}_{-174.51}$  &    $144.96^{17.3}_{-51.06}$ \\  IRAS06206-6315 & 0.09244 &    1.38  &    $12.13^{2.21}_{-5.39}$  &    $3.22^{0.89}_{-0.54}$  &    $78.51^{16.17}_{-30.69}$  &    $213.4^{13.74}_{-9.43}$ \\  IRAS07598+6508 & 0.1483 &    0.22  &    $1.0^{1.7}_{-0.59}$  &    $4.98^{5.59}_{-2.37}$  &    $96.56^{170.51}_{-60.04}$  &    $74.93^{13.3}_{-10.46}$ \\  IRAS08311-2459 & 0.1004 &    0.74  &    $1.18^{0.84}_{-0.49}$  &    $5.07^{2.71}_{-1.28}$  &    $165.17^{128.78}_{-60.21}$  &    $69.34^{11.08}_{-10.11}$ \\  IRAS08572+3915 & 0.05835 &    1.67  &    $0.49^{0.36}_{-0.15}$  &    $10.43^{3.39}_{-2.12}$  &    $443.13^{36.46}_{-43.89}$  &    $95.82^{8.98}_{-11.97}$ \\  IRAS09022-3615 & 0.05964 &    0.94  &    $0.67^{0.39}_{-0.28}$  &    $3.34^{1.99}_{-0.96}$  &    $164.3^{9.07}_{-7.83}$  &    $95.28^{11.97}_{-9.54}$ \\  IRAS10378+1109 & 0.1362 &    1.65  &    $1.14^{1.57}_{-0.45}$  &    $3.91^{1.86}_{-1.08}$  &    $182.55^{33.44}_{-6.78}$  &    $121.63^{5.69}_{-7.35}$ \\  IRAS10565+2448 & 0.043 &    1.69  &    $0.47^{0.41}_{-0.18}$  &    $5.68^{1.36}_{-1.68}$  &    $85.35^{16.82}_{-14.77}$  &    $142.66^{3.79}_{-3.43}$ \\  IRAS11095-0238 & 0.1066 &    1.22  &    $1.34^{5.09}_{-0.82}$  &    $3.54^{1.66}_{-1.67}$  &    $390.54^{211.33}_{-160.73}$  &    $140.62^{23.68}_{-12.75}$ \\  IRAS12071-0444 & 0.1284 &    1.67  &    $4.26^{1.58}_{-1.35}$  &    $1.55^{0.25}_{-0.44}$  &    $228.69^{95.08}_{-13.62}$  &    $142.81^{15.37}_{-19.74}$ \\  IRAS13120-5453 & 0.03076 &    2.08  &    $2.48^{2.31}_{-1.49}$  &    $4.46^{1.95}_{-1.22}$  &    $184.21^{140.39}_{-89.49}$  &    $197.84^{1.39}_{-5.12}$ \\  IRAS13451+1232 & 0.1217 &    1.84  &    $8.19^{1.21}_{-0.61}$  &    $1.67^{0.08}_{-0.21}$  &    $322.61^{35.84}_{-67.3}$  &    $130.6^{4.76}_{-8.16}$ \\  IRAS14348-1447 & 0.0827 &    2.24  &    $0.73^{1.26}_{-0.31}$  &    $3.77^{1.1}_{-1.0}$  &    $209.25^{70.78}_{-90.79}$  &    $199.34^{11.04}_{-2.2}$ \\  IRAS14378-3651 & 0.0676 &    2.65  &    $0.83^{6.02}_{-0.54}$  &    $3.38^{2.09}_{-0.94}$  &    $114.47^{82.42}_{-26.81}$  &    $198.56^{26.58}_{-3.49}$ \\  IRAS15250+3609 & 0.05516 &    2.31  &    $0.5^{0.94}_{-0.3}$  &    $2.1^{0.56}_{-0.34}$  &    $650.13^{69.35}_{-44.14}$  &    $153.97^{8.23}_{-7.16}$ \\  IRAS15462-0450 & 0.09979 &    1.39  &    $0.27^{0.43}_{-0.16}$  &    $2.91^{1.1}_{-1.26}$  &    $187.39^{161.35}_{-53.98}$  &    $166.57^{16.86}_{-34.6}$ \\  IRAS16090-0139 & 0.13358 &    1.01  &    $1.7^{2.62}_{-1.39}$  &    $4.42^{1.19}_{-1.49}$  &    $305.63^{326.12}_{-265.3}$  &    $136.43^{5.64}_{-5.3}$ \\  IRAS17208-0014 & 0.0428 &    2.55  &    $0.18^{0.02}_{-0.04}$  &    $1.14^{0.22}_{-0.01}$  &    $166.87^{66.33}_{-10.54}$  &    $198.93^{6.62}_{-0.24}$ \\  IRAS19254-7245 & 0.061709 &    1.6  &    $8.24^{2.27}_{-1.21}$  &    $4.23^{2.07}_{-1.2}$  &    $154.04^{5.2}_{-11.84}$  &    $95.37^{25.1}_{-21.62}$ \\  IRAS19297-0406 & 0.08573 &    2.21  &    $0.28^{0.41}_{-0.14}$  &    $3.82^{1.7}_{-1.28}$  &    $70.94^{68.06}_{-38.8}$  &    $195.56^{2.9}_{-5.88}$ \\  IRAS20087-0308 & 0.10567 &    1.5  &    $2.54^{1.06}_{-1.19}$  &    $1.9^{1.19}_{-0.54}$  &    $67.56^{48.77}_{-29.36}$  &    $196.41^{2.76}_{-3.16}$ \\  IRAS20100-4156 & 0.12958 &    2.54  &    $1.58^{3.11}_{-0.96}$  &    $4.19^{1.38}_{-1.47}$  &    $335.05^{104.17}_{-60.68}$  &    $198.37^{5.41}_{-7.53}$ \\  IRAS20414-1651 & 0.087084 &    2.96  &    $0.14^{0.04}_{-0.02}$  &    $9.49^{3.49}_{-0.72}$  &    $165.1^{31.41}_{-9.71}$  &    $194.78^{4.1}_{-8.12}$ \\  IRAS20551-4250 & 0.042996 &    1.77  &    $2.03^{0.6}_{-0.34}$  &    $2.83^{1.25}_{-0.69}$  &    $636.41^{49.67}_{-44.28}$  &    $147.0^{11.2}_{-10.93}$ \\  IRAS22491-1808 & 0.0778 &    2.01  &    $4.23^{0.22}_{-3.36}$  &    $1.89^{0.71}_{-0.23}$  &    $734.86^{25.11}_{-157.92}$  &    $223.09^{15.12}_{-17.16}$ \\  IRAS23128-5919 & 0.0446 &    1.33  &    $1.87^{1.86}_{-0.98}$  &    $4.61^{0.48}_{-0.64}$  &    $698.76^{65.73}_{-111.18}$  &    $227.74^{13.42}_{-7.57}$ \\  IRAS23230-6926 & 0.10659 &    1.98  &    $7.75^{0.65}_{-1.19}$  &    $4.4^{1.88}_{-1.11}$  &    $478.09^{39.74}_{-69.79}$  &    $218.79^{15.4}_{-15.51}$ \\  IRAS23253-5415 & 0.13 &    2.13  &    $0.18^{0.07}_{-0.04}$  &    $1.52^{0.33}_{-0.26}$  &    $149.77^{65.7}_{-51.19}$  &    $242.16^{3.92}_{-10.16}$ \\  IRAS23365+3604 & 0.0645 &    3.09  &    $5.63^{0.94}_{-4.44}$  &    $2.67^{0.07}_{-0.18}$  &    $210.06^{54.53}_{-36.89}$  &    $229.65^{8.56}_{-6.56}$ \\  Mrk1014 & 0.16311 &    2.62  &    $1.73^{2.48}_{-0.52}$  &    $3.57^{4.03}_{-1.21}$  &    $227.58^{168.96}_{-129.66}$  &    $52.81^{4.86}_{-0.99}$ \\  UGC5101 & 0.039367 &    2.93  &    $13.93^{0.5}_{-0.54}$  &    $1.37^{0.1}_{-0.12}$  &    $196.75^{3.92}_{-5.63}$  &    $196.66^{4.3}_{-5.78}$ \\  Mrk231 & 0.04217 &    0.58  &    $0.57^{0.5}_{-0.41}$  &    $2.31^{1.14}_{-0.79}$  &    $514.11^{185.21}_{-157.76}$  &    $145.48^{36.66}_{-14.34}$ \\  Mrk273 & 0.03778 &    1.63  &    $0.95^{0.38}_{-0.26}$  &    $5.02^{0.77}_{-1.21}$  &    $175.69^{4.72}_{-8.87}$  &    $135.74^{15.82}_{-5.76}$ \\  Mrk463 & 0.050355 &    1.5  &    $1.32^{0.36}_{-0.45}$  &    $3.4^{4.09}_{-0.92}$  &    $116.4^{67.27}_{-21.8}$  &    $116.17^{16.08}_{-11.1}$ \\  Arp220 & 0.018 &    5.23  &    $0.12^{0.03}_{-0.02}$  &    $1.83^{0.53}_{-0.39}$  &    $124.13^{1.71}_{-0.22}$  &    $246.08^{2.19}_{-1.06}$ \\  NGC6240 & 0.0244 &    1.62  &    $3.23^{0.8}_{-1.16}$  &    $3.96^{0.6}_{-0.24}$  &    $247.7^{19.01}_{-37.22}$  &    $76.64^{6.67}_{-6.98}$ \\   \hline
 \end{tabular}}
\end{table*}

\begin{table*}
	\centering
\begin{flushleft}
\textbf{Table A2.} Other fitted parameters for the galaxies in the HERUS sample
\end{flushleft}
	\label{tab:fittedB}
 \resizebox{\textwidth}{!}{\begin{tabular}{ lcccccccc }
 \hline 
Name & $t_{*} \ (10^7yr)$ & $\tau_{*} \ (10^7yr)$ & $r_2/r_1$ & $\tau_{uv}$ & $\theta_o \ (\degr)$ & $\theta_i \ (\degr)$ & $T_p$ \\ 
 \hline IRAS00188-0856 &    $3.22^{0.08}_{-0.19}$  &    $1.55^{0.1}_{-0.05}$  &    $57.42^{30.87}_{-21.08}$  &    $394.96^{77.68}_{-18.89}$  &    $46.42^{2.66}_{-3.95}$  &    $76.46^{4.46}_{-0.37}$ & - \\  IRAS00397-1312 &    $0.77^{0.14}_{-0.16}$  &    $2.01^{0.22}_{-0.23}$  &    $89.46^{6.71}_{-5.84}$  &    $329.52^{9.58}_{-9.24}$  &    $73.81^{0.27}_{-0.09}$  &    $83.12^{1.11}_{-0.87}$ & - \\  IRAS01003-2238 &    $1.38^{0.25}_{-0.19}$  &    $1.45^{0.28}_{-0.11}$  &    $52.48^{5.82}_{-3.25}$  &    $1310.68^{86.96}_{-24.24}$  &    $61.37^{0.19}_{-0.09}$  &    $69.32^{0.27}_{-0.2}$ & - \\  IRAS03158+4227 &    $3.03^{0.25}_{-0.39}$  &    $1.39^{0.15}_{-0.15}$  &    $30.93^{5.92}_{-3.6}$  &    $959.57^{33.8}_{-51.93}$  &    $53.97^{4.84}_{-2.01}$  &    $76.88^{0.62}_{-0.6}$ & - \\  IRAS03521+0028 &    $3.4^{0.07}_{-0.05}$  &    $1.55^{0.2}_{-0.2}$  &    $45.37^{1.6}_{-3.74}$  &    $1245.22^{81.44}_{-99.23}$  &    $58.9^{3.52}_{-3.74}$  &    $70.52^{1.82}_{-1.4}$ & - \\  IRAS05189-2524 &    $2.15^{0.42}_{-0.28}$  &    $2.09^{0.21}_{-0.37}$  &    $38.24^{4.1}_{-4.16}$  &    $911.92^{72.26}_{-33.33}$  &    $52.28^{3.43}_{-1.28}$  &    $63.61^{1.75}_{-0.6}$ &     $1050.6^{28.7}_{-23.2}$ \\  IRAS06035-7102 &    $0.84^{1.99}_{-0.14}$  &    $2.03^{0.12}_{-0.43}$  &    $43.38^{9.55}_{-4.72}$  &    $361.74^{106.82}_{-22.29}$  &    $48.28^{1.82}_{-8.36}$  &    $68.39^{2.01}_{-2.32}$ & - \\  IRAS06206-6315 &    $3.09^{0.2}_{-0.19}$  &    $1.38^{0.11}_{-0.11}$  &    $48.27^{4.85}_{-5.79}$  &    $695.61^{36.17}_{-84.44}$  &    $45.32^{4.21}_{-3.83}$  &    $70.29^{1.12}_{-1.09}$ & - \\  IRAS07598+6508 &    $2.83^{0.2}_{-0.52}$  &    $2.08^{0.47}_{-0.4}$  &    $36.43^{16.48}_{-10.28}$  &    $723.81^{145.32}_{-106.6}$  &    $63.02^{2.3}_{-1.76}$  &    $15.27^{2.16}_{-2.84}$ &     $1094.2^{72.0}_{-98.8}$ \\  IRAS08311-2459 &    $2.84^{0.15}_{-0.12}$  &    $1.22^{0.23}_{-0.11}$  &    $30.23^{6.94}_{-3.93}$  &    $650.82^{97.1}_{-82.22}$  &    $48.54^{4.38}_{-3.8}$  &    $60.15^{2.95}_{-1.86}$ & - \\  IRAS08572+3915 &    $0.64^{0.28}_{-0.1}$  &    $1.79^{0.34}_{-0.25}$  &    $58.5^{9.8}_{-8.38}$  &    $612.99^{14.44}_{-21.13}$  &    $72.66^{1.05}_{-0.89}$  &    $79.04^{0.74}_{-0.65}$ & - \\  IRAS09022-3615 &    $3.08^{0.24}_{-0.2}$  &    $1.41^{0.12}_{-0.25}$  &    $40.44^{8.88}_{-7.7}$  &    $308.95^{38.16}_{-30.87}$  &    $37.15^{3.16}_{-8.14}$  &    $63.27^{2.77}_{-3.59}$ & - \\  IRAS10378+1109 &    $2.98^{0.3}_{-0.22}$  &    $1.3^{0.35}_{-0.14}$  &    $31.19^{7.18}_{-5.31}$  &    $1427.8^{51.92}_{-135.04}$  &    $51.29^{1.97}_{-2.56}$  &    $72.63^{0.81}_{-2.14}$ & - \\  IRAS10565+2448 &    $3.47^{0.02}_{-0.02}$  &    $1.05^{0.05}_{-0.04}$  &    $44.7^{5.66}_{-5.41}$  &    $825.07^{38.72}_{-59.02}$  &    $54.12^{2.16}_{-1.54}$  &    $66.1^{0.6}_{-0.75}$ & - \\  IRAS11095-0238 &    $2.46^{0.81}_{-1.25}$  &    $1.92^{0.57}_{-0.38}$  &    $59.5^{11.12}_{-16.98}$  &    $904.31^{53.4}_{-37.8}$  &    $60.14^{2.46}_{-3.65}$  &    $76.33^{0.79}_{-0.89}$ & - \\  IRAS12071-0444 &    $0.63^{0.31}_{-0.12}$  &    $1.88^{0.02}_{-0.1}$  &    $38.36^{0.69}_{-2.4}$  &    $794.25^{24.03}_{-30.22}$  &    $47.29^{0.2}_{-3.41}$  &    $66.95^{0.48}_{-0.8}$ & - \\  IRAS13120-5453 &    $3.38^{0.08}_{-0.06}$  &    $1.05^{0.06}_{-0.02}$  &    $51.96^{3.76}_{-5.55}$  &    $554.66^{9.27}_{-41.27}$  &    $43.12^{5.92}_{-1.56}$  &    $67.39^{0.75}_{-2.69}$ & - \\  IRAS13451+1232 &    $1.49^{0.03}_{-0.15}$  &    $1.49^{0.15}_{-0.17}$  &    $39.07^{5.08}_{-3.41}$  &    $1456.38^{19.18}_{-42.13}$  &    $51.33^{0.54}_{-2.2}$  &    $64.28^{0.35}_{-1.55}$ &     $938.9^{33.3}_{-13.2}$ \\  IRAS14348-1447 &    $3.39^{0.07}_{-0.16}$  &    $2.16^{0.28}_{-0.23}$  &    $25.75^{3.24}_{-2.71}$  &    $802.79^{96.15}_{-140.11}$  &    $51.38^{4.0}_{-3.78}$  &    $70.64^{1.35}_{-1.71}$ & - \\  IRAS14378-3651 &    $3.43^{0.04}_{-0.03}$  &    $2.26^{0.23}_{-0.58}$  &    $43.77^{14.34}_{-4.42}$  &    $1035.7^{252.4}_{-114.91}$  &    $47.17^{4.29}_{-3.19}$  &    $68.96^{1.95}_{-0.61}$ & - \\  IRAS15250+3609 &    $2.99^{0.36}_{-0.64}$  &    $2.87^{0.32}_{-0.4}$  &    $34.51^{2.53}_{-2.87}$  &    $927.48^{26.0}_{-29.82}$  &    $55.78^{1.06}_{-1.94}$  &    $76.97^{0.82}_{-0.49}$ & - \\  IRAS15462-0450 &    $1.87^{0.59}_{-0.55}$  &    $2.25^{0.52}_{-0.32}$  &    $38.77^{17.9}_{-6.05}$  &    $532.12^{78.8}_{-55.67}$  &    $55.53^{7.7}_{-1.2}$  &    $60.96^{3.81}_{-1.62}$ & - \\  IRAS16090-0139 &    $3.44^{0.04}_{-0.08}$  &    $2.01^{0.25}_{-0.31}$  &    $46.52^{4.75}_{-3.4}$  &    $545.12^{49.17}_{-24.69}$  &    $46.62^{6.14}_{-3.44}$  &    $72.82^{1.55}_{-1.13}$ & - \\  IRAS17208-0014 &    $3.44^{0.0}_{-0.04}$  &    $1.32^{0.01}_{-0.03}$  &    $46.3^{2.02}_{-1.23}$  &    $633.32^{1.29}_{-16.95}$  &    $48.08^{3.57}_{-2.53}$  &    $65.94^{0.29}_{-0.44}$ & - \\  IRAS19254-7245 &    $1.13^{0.13}_{-0.2}$  &    $2.53^{0.42}_{-0.45}$  &    $43.82^{7.91}_{-4.61}$  &    $780.52^{42.86}_{-98.51}$  &    $55.46^{2.31}_{-1.32}$  &    $68.66^{0.75}_{-0.66}$ & - \\  IRAS19297-0406 &    $3.44^{0.01}_{-0.06}$  &    $2.0^{0.22}_{-0.5}$  &    $78.41^{9.95}_{-23.53}$  &    $824.04^{81.58}_{-40.58}$  &    $41.04^{2.88}_{-4.01}$  &    $64.06^{2.21}_{-1.97}$ & - \\  IRAS20087-0308 &    $3.48^{0.01}_{-0.02}$  &    $1.04^{0.04}_{-0.03}$  &    $49.76^{7.82}_{-4.88}$  &    $316.83^{15.92}_{-16.97}$  &    $58.14^{1.66}_{-2.27}$  &    $80.16^{1.72}_{-2.36}$ & - \\  IRAS20100-4156 &    $2.0^{0.45}_{-0.9}$  &    $2.54^{0.68}_{-0.72}$  &    $45.69^{9.01}_{-5.02}$  &    $613.02^{65.63}_{-35.23}$  &    $47.39^{3.11}_{-2.75}$  &    $73.59^{0.81}_{-0.97}$ & - \\  IRAS20414-1651 &    $3.4^{0.04}_{-0.03}$  &    $1.22^{0.01}_{-0.04}$  &    $51.91^{4.43}_{-5.34}$  &    $620.3^{90.57}_{-47.04}$  &    $62.2^{0.76}_{-2.88}$  &    $70.79^{0.63}_{-0.31}$ & - \\  IRAS20551-4250 &    $1.26^{0.36}_{-0.22}$  &    $1.7^{0.37}_{-0.49}$  &    $42.29^{4.32}_{-4.7}$  &    $732.36^{24.46}_{-17.98}$  &    $48.65^{2.84}_{-2.31}$  &    $74.11^{0.85}_{-1.19}$ & - \\  IRAS22491-1808 &    $1.46^{1.79}_{-0.48}$  &    $2.45^{0.22}_{-0.3}$  &    $32.91^{4.86}_{-4.48}$  &    $1237.53^{109.38}_{-59.62}$  &    $49.02^{1.98}_{-1.38}$  &    $71.39^{0.67}_{-1.71}$ & - \\  IRAS23128-5919 &    $1.55^{0.41}_{-0.5}$  &    $1.21^{0.1}_{-0.13}$  &    $31.21^{3.91}_{-4.0}$  &    $612.96^{34.8}_{-35.47}$  &    $44.41^{2.01}_{-1.94}$  &    $63.37^{0.46}_{-0.81}$ & - \\  IRAS23230-6926 &    $1.32^{0.27}_{-0.32}$  &    $1.58^{0.28}_{-0.19}$  &    $45.45^{5.7}_{-4.53}$  &    $813.11^{29.03}_{-36.16}$  &    $48.57^{2.3}_{-2.03}$  &    $74.03^{0.72}_{-0.7}$ & - \\  IRAS23253-5415 &    $3.21^{0.17}_{-0.1}$  &    $1.39^{0.45}_{-0.24}$  &    $93.24^{2.3}_{-4.69}$  &    $1378.25^{41.64}_{-26.05}$  &    $49.08^{1.76}_{-1.71}$  &    $68.78^{0.59}_{-0.98}$ & - \\  IRAS23365+3604 &    $2.54^{0.1}_{-0.18}$  &    $1.19^{0.22}_{-0.04}$  &    $26.41^{0.51}_{-1.68}$  &    $1442.99^{22.88}_{-19.41}$  &    $51.7^{2.1}_{-3.48}$  &    $74.55^{0.19}_{-0.36}$ & - \\  Mrk1014 &    $2.13^{0.2}_{-0.14}$  &    $1.18^{0.16}_{-0.13}$  &    $61.14^{11.1}_{-9.45}$  &    $286.33^{103.28}_{-19.2}$  &    $59.76^{1.04}_{-0.85}$  &    $18.06^{1.83}_{-1.22}$ & - \\  UGC5101 &    $1.05^{0.09}_{-0.06}$  &    $1.51^{0.2}_{-0.08}$  &    $76.09^{4.28}_{-8.96}$  &    $339.69^{22.83}_{-23.61}$  &    $46.29^{1.8}_{-2.13}$  &    $75.48^{0.22}_{-1.19}$ & - \\  Mrk231 &    $2.1^{0.33}_{-0.36}$  &    $1.52^{0.14}_{-0.12}$  &    $40.1^{4.43}_{-6.97}$  &    $636.02^{35.16}_{-22.87}$  &    $54.24^{2.76}_{-3.08}$  &    $63.79^{1.22}_{-1.03}$ & - \\  Mrk273 &    $2.98^{0.12}_{-0.16}$  &    $1.33^{0.12}_{-0.08}$  &    $37.9^{4.06}_{-1.1}$  &    $788.22^{106.94}_{-91.5}$  &    $44.88^{1.48}_{-2.48}$  &    $66.61^{0.84}_{-1.83}$ & - \\  Mrk463 &    $1.67^{0.46}_{-0.38}$  &    $2.18^{0.16}_{-0.21}$  &    $46.87^{5.6}_{-18.56}$  &    $517.08^{68.32}_{-100.63}$  &    $63.17^{3.89}_{-1.71}$  &    $63.26^{1.56}_{-3.02}$ & - \\  Arp220 &    $2.95^{0.04}_{-0.02}$  &    $1.11^{0.08}_{-0.07}$  &    $59.58^{15.48}_{-7.3}$  &    $1447.17^{16.31}_{-4.36}$  &    $48.24^{0.27}_{-1.42}$  &    $78.29^{0.3}_{-0.57}$ & - \\  NGC6240 &    $3.23^{0.18}_{-0.19}$  &    $1.25^{0.23}_{-0.16}$  &    $44.07^{4.56}_{-7.74}$  &    $1097.07^{95.22}_{-76.14}$  &    $54.62^{3.61}_{-2.12}$  &    $71.3^{0.66}_{-0.67}$ & - \\     \hline 
 \end{tabular}}
\end{table*}

\begin{table*}
	\centering
\begin{flushleft} 
\textbf{Table A3.} Selected extracted physical quantities for IRAS~08572+3915. For the AGN and total luminosities the anisotropy-corrected luminosities are given.
\end{flushleft} 
	\label{tab:3915phy1}
 \resizebox{\textwidth}{!}{\begin{tabular}{ lcccccccc }
 \hline 
AGN model & $L_{AGN}^c$ & $L_{SB}$ & $L_{sph}$ & $L_{tot}^c$ & $\dot{M}^{age}_{*}$ & $\dot{M}_{sph}$ \\ 
     &     $10^{12} L_\odot$ & $10^{11} L_\odot$ & $10^{10} L_\odot$  & $10^{12} L_\odot$ & $M_\odot yr^{-1}$ & $M_\odot yr^{-1}$ \\  
 \hline CYGNUS &    $12.85^{1.1}_{-1.36}$  &    $5.46^{0.42}_{-1.07}$  &    $1.67^{0.08}_{-0.09}$  &    $13.34^{1.21}_{-1.4}$  &    $154.0^{17.93}_{-48.26}$  &    $1.52^{0.15}_{-0.13}$ \\  
                   &            &  &   \\ 
 \cite{fritz06} &    $4.14^{0.0}_{-0.55}$  &    $2.07^{3.35}_{-0.0}$  &    $2.73^{0.0}_{-1.0}$  &    $4.38^{0.0}_{-0.22}$  &    $55.93^{75.42}_{-0.0}$  &    $2.31^{0.0}_{-1.56}$ \\ 
                   &            &  &   \\ 
 SKIRTOR &    $0.0^{0.01}_{-0.0}$  &    $14.35^{0.15}_{-0.21}$  &    $1.59^{0.1}_{-0.03}$  &    $1.45^{0.02}_{-0.02}$  &    $428.1^{9.09}_{-8.14}$  &    $1.46^{0.1}_{-0.07}$ \\ 
                   &            &  &   \\ 
 \cite{sieb15} &   $0.01^{0.01}_{-0.0}$  &    $14.33^{0.31}_{-0.22}$  &    $1.57^{0.05}_{-0.05}$  &    $1.46^{0.03}_{-0.03}$  &    $430.1^{6.11}_{-10.15}$  &    $1.42^{0.1}_{-0.07}$ \\  \hline 
 \end{tabular}}
\end{table*}

\begin{table*}
	\centering
\begin{flushleft} 
\textbf{Table A4.} Other extracted physical quantities for IRAS~08572+3915
\end{flushleft}
	\label{tab:3915phy2}
 \resizebox{\textwidth}{!}{\begin{tabular}{ lcccccccc }
 \hline 
AGN model & $\dot{M}_{tot}$ & ${M}^{*}_{sph}$ & ${M}^{*}_{SB}$ & ${M}^{*}_{tot}$& $F_{AGN}$ & $A$ \\ 
     &     $M_\odot yr^{-1}$ & $10^{10} M_\odot$ & $10^{9} M_\odot$ & $10^{10} M_\odot$ & $   $ & $   $ \\  
 \hline CYGNUS &    $155.5^{18.02}_{-48.19}$  &    $2.19^{0.19}_{-0.15}$  &    $0.9^{0.14}_{-0.07}$  &    $2.28^{0.22}_{-0.16}$  &     $0.96^{0.01}_{-0.01}$  &     $14.2^{1.38}_{-1.29}$ \\  
                    &            &  &   \\ 
 \cite{fritz06} &    $58.24^{73.86}_{-0.0}$  &    $2.1^{2.28}_{-0.0}$  &    $1.63^{0.18}_{-0.0}$  &    $2.27^{2.3}_{-0.0}$  &     $0.95^{0.0}_{-0.08}$  &     $3.53^{0.5}_{-0.0}$ \\ 
                    &            &  &   \\ 
 SKIRTOR &    $429.8^{8.91}_{-8.48}$  &    $2.24^{0.18}_{-0.13}$  &    $2.11^{0.03}_{-0.03}$  &    $2.45^{0.18}_{-0.14}$  &     $0.0^{0.0}_{-0.0}$  &     $1.4^{1.14}_{-0.36}$ \\ 
                    &            &  &   \\ 
 \cite{sieb15} &    $431.5^{6.05}_{-10.09}$  &    $2.18^{0.14}_{-0.13}$  &    $2.1^{0.07}_{-0.04}$  &    $2.4^{0.14}_{-0.14}$  &     $0.0^{0.0}_{-0.0}$  &     $3.64^{0.38}_{-0.66}$ \\   \hline 
 \end{tabular}}
\end{table*}

\begin{table*}
	\centering
\begin{flushleft} 
\textbf{Table A5.} Selected extracted physical quantities for VV~340a. For the AGN and total luminosities the anisotropy-corrected luminosities are given.
\end{flushleft}
	\label{tab:vv340aphy1}
 \resizebox{\textwidth}{!}{\begin{tabular}{ lccccccc }
 \hline 
Host galaxy model  & $L_{AGN}^c$ & $L_{SB}$ &  $L_{sph}$ or $L_{disc}$ & $L_{tot}^c$ & $\dot{M}^{age}_{*}$ & $\dot{M}_{sph}$ \\  
$  $ & $10^{12} L_\odot$ & $10^{11} L_\odot$ &  $10^{10} L_\odot$ &  $10^{12} L_\odot$ &  $M_\odot yr^{-1}$ & $M_\odot yr^{-1}$  \\ \hline
 Spheroidal &  $0.05^{0.01}_{-0.02}$  &    $2.55^{0.33}_{-0.45}$  &    $16.12^{1.71}_{-0.55}$  &    $0.46^{0.05}_{-0.04}$  &    $78.21^{36.21}_{-3.35}$  &    $3.81^{1.09}_{-0.62}$ \\ 
                    &            &  &   \\ 
 Disc &    $0.25^{0.06}_{-0.04}$  &    $1.48^{0.23}_{-0.33}$  &    $30.98^{2.1}_{-0.67}$  &    $0.72^{0.04}_{-0.03}$  &     $35.41^{5.26}_{-8.42}$  &     $11.56^{1.28}_{-0.29}$  \\   \hline 
 \end{tabular}}
\end{table*}

\begin{table*}
	\centering
\begin{flushleft} 
\textbf{Table A6.} Other extracted physical quantities for VV~340a
\end{flushleft} 
	\label{tab:vv340aphy2}
 \resizebox{\textwidth}{!}{\begin{tabular}{ lccccccc }
 \hline 
Host galaxy model &  $\dot{M}_{tot}$ &  ${M}^{*}_{sph}$ or ${M}^{*}_{disc}$  & ${M}^{*}_{SB}$ & ${M}^{*}_{tot}$& $F_{AGN}$ & $A$ \\  
$  $ & $M_\odot yr^{-1}$ & $10^{10} M_\odot$ & $10^{9} M_\odot$ & $10^{10} M_\odot$ & $   $ & $   $ \\  \hline
 Spheroidal & $81.98^{36.33}_{-3.32}$  &    $21.24^{3.99}_{-1.01}$  &    $2.38^{0.82}_{-0.13}$  &    $21.47^{4.04}_{-1.0}$  &     $0.1^{0.02}_{-0.03}$  &     $1.65^{0.36}_{-0.67}$ \\ 
                    &            &  &   \\ 
 Disc &  $47.79^{4.43}_{-0.29}$  &    $21.34^{0.78}_{-0.49}$  &    $0.36^{0.03}_{-0.07}$  &    $21.38^{0.77}_{-0.48}$  &     $0.35^{0.07}_{-0.04}$  &     $8.3^{0.84}_{-0.8}$ \\  \hline 
 \end{tabular}}
\end{table*}

\begin{table*}
	\centering
\begin{flushleft} 
\textbf{Table A7.} Selected extracted physical quantities for HELP\_J100156.75+022344.7. For the AGN and total luminosities the anisotropy-corrected luminosities are given.
\end{flushleft} 
	\label{tab:helpphy1}
 \resizebox{\textwidth}{!}{\begin{tabular}{ lcccccccc }
 \hline 
AGN model & $L_{AGN}^c$ & $L_{SB}$ & $L_{sph}$ & $L_{tot}^c$ & $\dot{M}^{age}_{*}$ & $\dot{M}_{sph}$ \\  
     &      $10^{12} L_\odot$ & $10^{11} L_\odot$ & $10^{10} L_\odot$ & $10^{12} L_\odot$ & $M_\odot yr^{-1}$ & $M_\odot yr^{-1}$ \\ 
 \hline CYGNUS &    $23.54^{0.24}_{-5.08}$  &    $40.83^{4.8}_{-5.05}$  &    $9.88^{35.02}_{-1.78}$  &    $27.45^{0.28}_{-4.27}$  &    $994.3^{86.46}_{-110.2}$  &    $7.29^{27.14}_{-2.58}$ \\  
                    &            &  &   \\ 
 \cite{fritz06} &    $14.62^{1.28}_{-1.54}$  &    $55.31^{7.92}_{-7.06}$  &    $9.7^{7.53}_{-1.59}$  &    $20.27^{1.15}_{-1.06}$  &    $1349.0^{160.7}_{-204.4}$  &    $8.19^{4.92}_{-1.19}$ \\  
                    &            &  &   \\ 
 SKIRTOR &    $15.29^{4.65}_{-1.91}$  &    $50.07^{12.52}_{-6.45}$  &    $10.62^{4.41}_{-2.96}$  &    $20.45^{5.35}_{-2.02}$  &    $1267.0^{285.3}_{-206.3}$  &    $8.2^{3.17}_{-2.2}$ \\ 
                    &            &  &   \\ 
 \cite{sieb15} &   $26.58^{5.32}_{-1.89}$  &    $2.27^{17.66}_{-2.22}$  &    $9.71^{6.72}_{-1.57}$  &    $28.83^{3.34}_{-3.78}$  &    $54.48^{461.9}_{-53.2}$  &    $8.35^{4.4}_{-1.89}$ \\  \hline 
 \end{tabular}}
\end{table*}

\begin{table*}
	\centering
\begin{flushleft} 
\textbf{Table A8.} Other extracted physical quantities for HELP\_J100156.75+022344.7
\end{flushleft} 
	\label{tab:helpphy2}
 \resizebox{\textwidth}{!}{\begin{tabular}{ lcccccccc }
 \hline 
AGN model & $\dot{M}_{tot}$ & ${M}^{*}_{sph}$ & ${M}^{*}_{SB}$ & ${M}^{*}_{tot}$& $F_{AGN}$ & $A$ \\ 
     &     $M_\odot yr^{-1}$ & $10^{10} M_\odot$ & $10^{9} M_\odot$ & $10^{10} M_\odot$ & $   $ & $   $ \\  
 \hline CYGNUS &    $999.0^{132.2}_{-108.1}$  &    $4.13^{0.31}_{-0.36}$  &    $17.57^{0.98}_{-5.36}$  &    $5.74^{0.23}_{-0.17}$  &     $0.85^{0.01}_{-0.05}$  &     $1.22^{0.07}_{-0.04}$ \\  
                    &            &  &   \\ 
 \cite{fritz06} &    $1357.0^{160.4}_{-196.8}$  &    $3.41^{0.46}_{-0.39}$  &    $18.32^{7.54}_{-5.16}$  &    $5.33^{0.34}_{-0.54}$  &     $0.72^{0.04}_{-0.03}$  &     $1.55^{0.07}_{-0.04}$ \\ 
                    &            &  &   \\ 
 SKIRTOR &    $1273.0^{292.2}_{-203.0}$  &    $3.71^{0.45}_{-0.27}$  &    $19.97^{14.84}_{-8.46}$  &    $5.66^{1.69}_{-0.89}$  &     $0.74^{0.05}_{-0.04}$  &     $1.45^{0.24}_{-0.12}$ \\ 
                    &            &  &   \\ 
 \cite{sieb15} &    $62.2^{464.2}_{-49.23}$  &    $3.16^{0.32}_{-0.53}$  &    $0.94^{9.71}_{-0.93}$  &    $3.18^{1.06}_{-0.21}$  &     $0.99^{0.0}_{-0.07}$  &     $1.36^{0.08}_{-0.2}$ \\   \hline 
 \end{tabular}}
\end{table*}

\clearpage

\label{lastpage}

\end{document}